\documentclass[final,3p,times,twocolumn]{elsarticle}

\usepackage{amssymb}
\usepackage{amsmath}
\usepackage{amsthm}
\usepackage{bm}
\usepackage{booktabs}
\usepackage{graphicx}
\usepackage{epsfig}
\usepackage{epstopdf}

\usepackage{textcomp,booktabs}
\usepackage[usenames,dvipsnames]{color}
\usepackage{colortbl}
\usepackage{threeparttable}

\definecolor{mygray}{gray}{.9}
\definecolor{mywhite}{rgb}{1,1,1}

\journal{}

\begin{document}

\begin{frontmatter}


\title{Universality of preference behaviors in online music-listener bipartite networks: \\ A Big Data analysis}

\author[a,b,c]{Xiao-Pu Han}
\ead{xp@hznu.edu.cn}
\author[c]{Fen Lin}
\author[c]{Jonathan J. H. Zhu}
\author[d]{Tarik Hadzibeganovic}
\ead{tarik.hadzibeganovic@gmail.com}

\address[a]{Alibaba Research Center for Complexity Sciences, Hangzhou Normal University, Hangzhou 311121, China}
\address[b]{Institute of Information Economy and Alibaba Business School, Hangzhou Normal University, Hangzhou 311121, China}
\address[c]{Department of Media and Communication, City University of Hong Kong, Hong Kong, China}
\address[d]{Institute of Psychology, Faculty of Natural Sciences, University of Graz, Graz 8010, Austria}


\begin{abstract}
We investigate the formation of musical preferences of millions of users of the NetEase Cloud Music (NCM), one of the largest online music platforms in China. We combine the methods from complex networks theory and information sciences within the context of Big Data analysis to unveil statistical patterns and community structures underlying the formation and evolution of musical preference behaviors. Our analyses address the decay patterns of music influence, users' sensitivity to music, age and gender differences, and their relationship to regional economic indicators. Employing community detection in user-music bipartite networks, we identified eight major cultural communities in the population of NCM users. Female users exhibited higher within-group variability in preference behavior than males, with a major transition occurring around the age of 25. Moreveor, the musical tastes and the preference diversity measures of women were also more strongly associated with economic factors. However, in spite of the highly variable popularity of music tracks and the identified cultural and demographic differences, we observed that the evolution of musical preferences over time followed a power-law-like decaying function, and that NCM listeners showed the highest sensitivity to music released in their adolescence, peaking at the age of 13. Our findings suggest the existence of universal properties in the formation of musical tastes but also their culture-specific relationship to demographic factors, with wide-ranging implications for community detection and recommendation system design in online music platforms.  

\end{abstract}

\begin{keyword}
Musical preferences \sep complex networks \sep community detection \sep cross-cultural universals \sep gender differences \sep Big Data.


\end{keyword}

\end{frontmatter}


\section{Introduction}
\label{}

As a typical universal cultural feature \cite{Parkinpress,Savage2015,Ravignani,Greenberg2022}, music plays an important and irreplaceable role in our everyday life \cite{levitin}. Music not only provides pleasure, but it also constructs and shapes our social fabric \cite{ball2008}. It has been used widely for emotional regulation \cite{juslin}, pain management \cite{bernatzky}, enhancement of cognitive abilities \cite{schellenberg}, or for support of coordinated movement behaviors such as dance \cite{large2000}. Music has evidently played a substantial role in the evolution of various cultural hallmarks, including the preservation of knowledge, development of rituals and religious behaviors, as well as in the emergence of group cohesion that has been 
vital for our survival \cite{levitin}. 

Due to its multidimensional nature, music can be studied from an array of different perspectives \cite{ball2008}. For instance, one line of research focuses on musical features such as rhythmic patterns \cite{coca2016}, their statistical properties \cite{nakamukaneko} and influences on music composition \cite{pachet}, and their effects on individual music appreciation \cite{ball2008,tighe,cowenetal}. Studies of this kind have spawned a series of fruitful investigations into the complexity of music \cite{levitinchordia} and its multi-leveled relationship to human emotions and behavior \cite{ball2008,cowenetal,suzhou,ballbook}. Another line of investigation emphasizes the socio-cultural significance of music \cite{Askin2017} and its developmental and evolutionary origins \cite{Ravignani,nakamukaneko,serraetal,trainor,merker,fitch,miton,bbssavage}. Individual musical preferences \cite{hargreaves2021} can be seen as a bridge connecting these two perspectives, rightfully attracting much recent attention across a number of disciplines \cite{Parkinpress,Rent2011,Bonneville2013,Greenberg2015,Lietal2018,Mellander2018,Park2016,Clauset,lxuetal,Fricke}. However, previous investigations of musical preferences have largely gravitated around the laboratory studies of individual music-listener behaviors in the offline world, whereas our understanding of musical tastes as they emerge and develop online, in the context of modern social networking technologies and music streaming platforms, remains largely limited.

In the present paper, we study the development of musical preferences in one of the largest online music communities in China. We combine the tools and methods from complex networks theory, information sciences, and physics within the context of Big Data analysis to unveil statistical universality patterns of musical preferences and to detect the underlying listenership communities that emerge in online social networks. We further provide quantitative characterization of a number of variables related to musical preferences in the studied population of online music listeners, 
including their tagging diversity, community diversity, individual and global musical preferences, and the users' overall sensitivity to music. We then model the obtained empirical data to identify relevant features of musical taste development, such as the age of users' peak sensitivity to popular music. We finally perform a series of analyses addressing the relationships between music preference variables and demographic factors such as gender, age, and economic development. 

By applying network models to large-scale user-behavior data from online music platforms, we show that global musical preferences can be captured and analyzed more effectively than what is usually attained in laboratory or field settings by assessing the habits and tastes of individuals or small groups of music listeners. In addition, we demonstrate that such network modeling and associated quantitative analyses of large-scale datasets can reveal statistical patterns and universal features of music identity formation, which would otherwise remain undetected through classical surveys or laboratory experiments. 

Our results are interpreted in terms of contemporary psychological and socio-economic theories of musical taste development, as well as through the lens of recent technological frameworks and complex systems approaches to user preference behavior. Our findings thus contribute to a more integrative and multidisciplinary understanding of the development of preference behaviors in complex social systems, which is relevant for a wide range of applications including user profiling and community detection in online music networks, measurement of online content popularity, and the design of playlist generation and music recommendation systems for online social media.

\subsection{Previous research and its limitations}

Employing personality questionnaires and surveys on musical tastes, a series of studies have previously examined individual differences in musical preference behaviors and the relationships between people's taste in music and socio-psychological factors \cite{Rent2011,Greenberg2015,Rawlings1997,Langmeyer2012}. Since personality and social behaviors tend to vary as a function of demographic factors, many recent studies have investigated 
age trends and the associated temporal stability in the development of musical preferences \cite{Bonneville2013,Bonneville2017a,BRust2017,BTuomas2017}. Moreover, demographic factors such as age, sex, or economic status, have recently been viewed as even more influential than psychological factors 
as predictors of genre-based musical preferences in laboratory offline settings \cite{Bonneville2013,Mellander2018}. 

Nevertheless, despite offering rich insights into the origins of musical preferences \cite{Rent2011,Greenberg2015}, many earlier investigations have suffered consistently from several conspicuous gaps. First, previous studies have focused mostly on verifying the existence of links between personality traits and musical preferences, in spite of repeated unsatisfactory results with rather small effect sizes (for a discussion, see e.g \cite{Greenberg2015}). Consequentially, these studies have often neglected other prominent aspects of musical preferences formation, such as the global patterns of music 
popularity \cite{Askin2017,kpopularity,guerreroetal}, their variability with demographic factors, and their universal features that may be shared across times and cultures. Survey-based methodologies or laboratory experiments are limited in approaching such questions, as for example, due to enormous costs, they are typically restricted to small sample sizes that are often not sufficiently representative of whole populations and may therefore conceal the underlying global trends. Critically, these small-sample studies are additionally limited by considering a rather narrow range of participants' ages, such as only adolescents and younger adults or only middle-aged adult music listeners (for a discussion on this issue, see Ref. \cite{Bonneville2013}). 

Secondly, the vast majority of previous studies on musical preferences considered only individual-level offline perspectives, whereas the direct investigation of musical tastes as they emerge and evolve in social contexts, especially in online social networks, is still lacking. Indeed, even though the study of online preference behavior has generally attracted much recent attention \cite{Guoetal2017,liuwang2018,arrigo21}, our understanding of users' {\it musical preferences} as they emerge and develop in online music-based communities is still largely limited \cite{Parkinpress,Lietal2018,Park2016,Clauset}. Nevertheless, a more systematic study of social networks in online music-streaming platforms and a deeper understanding of individual preferences in such online communities can be mutually revealing and highly informative for the design of novel music recommendation systems \cite{skleeetal2010} or for the measurement of popularity and influence of a wide variety of online contents \cite{Wangetal2016,gufandi,wangtianlu}.

\subsection{Methodological advances in Big Data analysis}

The rise of large-scale online communities and Big Data technology have enabled access to huge amounts of information. In recent years, the related large-scale 
data analyses have been addressing nearly all aspects of social networking, including user behavior, system design, performance analysis, or privacy issues, revealing a number of previously unknown trends and global patterns that remained largely undetected when employing classical data evaluation tools and techniques \cite{Askin2017,garciatanase,wangfujita,youetal2016,krakauerconst,scheffer,dedeopnas,subbubigdatares}. Within this context, especially in the study of music content popularity in online social media, the network-theoretic structural approach has gained considerable traction \cite{Askin2017}. Importantly, by combining Big Data and network theoretic analyses of preference behaviors in online music-based communities, we have the unique opportunity to study not only the large-scale picture of musical preferences formation, but also to understand how musical tastes are shaped by the psychological and social forces specifically emanating from the use of networking technologies and the associated computer-mediated interactions occurring among millions of users. 

Coupling large-scale dataset analysis with network science can additionally allow us to paint a more global picture of how various levels of sophistication of the online technology usage \cite{Pengzhu2011}, such as online skills, users' tagging behavior, or the diversity of online activities, contribute to the evolution of musical preferences. Excitingly, such studies will undoubtedly offer integrative insights into the combined effects of psychological, social, technological, and economic factors on musical taste formation and change.

\subsection{Research aims and hypotheses}

Since previous studies have mostly investigated tastes of Western music listeners (e.g. the US American users of Spotify, an online music streaming service), our present study focuses on the Big Data analysis of musical preferences of NetEase Cloud Music \cite{NCM}, one of the largest music streaming platforms and music-based online social networks in China. The first aim of this study was to uncover age trends and age-related global behavioral patterns in musical taste formation. By investigating these global features of musical preferences in China, and by comparing them to those observed with Western users of music-based social media, we further aimed at identifying general properties and laws of preference behaviors that may be shared universally across distinct cultures and epochs. Since online one-to-many interactions typically promote higher levels of cultural isolation and diversity \cite{Keijzer2018}, finding such cross-culturally shared behavioral patterns in online music-based communities would be a strong indicator of universality in the formation of musical preferences. 

Specifically, different from earlier small-sampled laboratory studies \cite{Holbrook1989}, and in line with recent large-scale analyses of online music streaming services in the US \cite{Stephens-Davidowitz2018}, we expected that the development of a listener's sensitivity to music in China would reach its peak prior to the early adulthood, in the early adolescence of an individual. Thus, the strongest influence on musical tastes of adult Chinese listeners should occur already in their early teens, representing the first major transition in the development of their musical preference behavior. We derive this hypothesis from interactionist theories (see e.g. \cite{Swann2002}), postulating that changes in an individual's development of musical preferences should be coupled to the associated changes in basic psychological needs and emotions, and to the emerging social identity formation of an 
individual (also see \cite{Rent2011,Bonneville2013}). 

Next, with respect to the temporal stability of online music influence, we expected to observe its slow-falling decay that would mimic patterns 
typically found in the time series of popular topics or in the outbreaks of memes in online social media \cite{Crane2008,Leskovec2009}. Furthermore, similar to earlier findings obtained with participants from the US and UK (e.g. \cite{Bonneville2013}), we predicted that while interest in some musical dimensions would decrease with age in Chinese music consumers, their preferences for other musical dimensions would increase in adulthood. 

Beyond this investigation of age trends in musical preferences, we addressed several specific assumptions related to gender differences in musical taste formation, their variability with age, and with regional economic development indicators. More specifically, we expected that during adolescence, relatively to their male counterparts, female listeners would be characterized by more heterogeneous music preference behaviors that would also be stronger associated with economic indicators. Thus, musical preferences and attitudes should reflect gender and economic divides \cite{Mellander2018}, particularly prior to adulthood of Chinese music listeners. However, in accordance with the social role theory \cite{Eagly2018}, we predicted that this gender gap in the adolescence should decrease with the onset of the early adulthood and with the increasing economic development, representing the second major transition in the development of musical preferences. In addition, employing network theoretic analysis and the associated community detection algorithm, we expected to identify distinct communities of NCM users that would allow us to further investigate culture-specific musical preference behaviors at the community and group levels. 

\section{Methods}

\subsection{Dataset}

\begin{figure}
\begin{center}
\hspace*{-0.5em}{
\includegraphics[width=8.75cm]{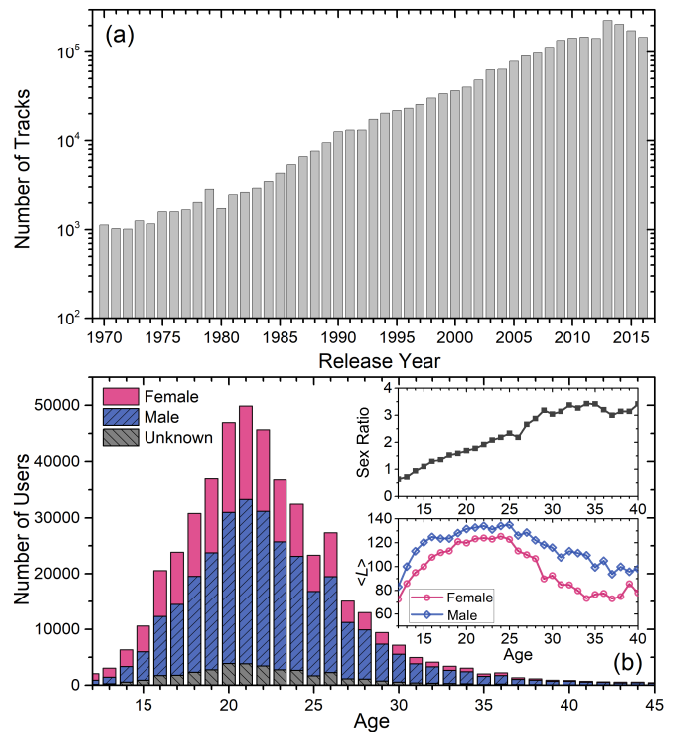}}
\end{center}
\vspace{-14pt}
\caption{(Color online)(a) The number of FP tracks as a function of the release year. (b) The distribution of users' age by gender. The numbers of male and female users by age are denoted by blue and pink bars, respectively. The upper inset shows the sex ratio of users (the number of male vs. female users) as a function of age, and the lower inset depicts the number of songs in users' FPs vs. age, separately for female and male users.
}\label{track_year}
\end{figure}

NetEase Cloud Music (http:$\verb|//|$music.163.com, hereafter abbreviated as NCM) is one of the world's largest music-based online social networks \cite{NCM}. Since its release in April 2013, hundreds of millions of users have registered with this music streaming platform. Using NCM software on a personal 
computer or smartphone, users can search, listen to, or download over 10 million songs from its online music library. Once registered with NCM, users can 
also build and edit playlists with their favorite songs, share their feelings and comments, or recommend music in a Twitter-like online section of NCM, or on other online social media. Here, we use the term ``songs" in a general sense to denote a piece of music with or without lyrics, and we further interchangeably use the terms ``song" and ``track" throughout this article.

The NCM dataset that was analyzed in the present study was collected from the API of NCM by web crawlers in the period from September 7 to September 27, 2016 \footnote{The dataset is freely available from the authors upon reasonable request.}. We selected to crawl the first 200,000 playlist entries from all available 2 million playlists. In total, we obtained a sample representing $10{\%}$ of all of the NCM's playlists, in which 30,562,590 playlists contained a total of 4,261,266 different tracks.

There were two types of playlists: One is called the Favorite Playlist (FP), which is created automatically when a user marks a track as a favorite for the first time. This list contains all the marked tracks of the user. Each user is allowed to create only one FP. The other type is the General Playlist, which can be freely created and modified by users. Each user can create no more than 1000 General Playlists. Our analysis mainly focuses on FPs, because they contain the music that a user listens to most frequently, thus reflecting the user's typical musical preferences. In total, we collected 4,499,164 FPs, in which 3,028,351 FPs corresponded to still active users, containing a total of 2,247,960 different songs. The remaining tracks come from General Playlists, amounting to 26,063,426 in total.

Each playlist contains detailed information about each track included in the playlist, such as a song's ID, title, its corresponding album, 
album release date, URL information, etc. As shown in Figure \ref{track_year}(a), users typically prefer new songs over old ones. Most favorite songs 
of NCM users were released in the last decade.

Each playlist also contains the information about its user, such as gender, age, city, and province. However, we were not able to access any information that could reveal user's personal identity. In our analyses, we thus only used the information that can be accessed publicly from NCM via web crawlers, and our data collection did not involve any interaction with or manipulation of any research subjects. 

Figure \ref{track_year}(b) shows the age and gender distributions of NCM users that were collected for the present study. We see that most users were in their 20s. However, we note here that age, as well as other information, was self-reported by users at the time of their account registration, 
and may thus not always be completely accurate. Considering the reliability of age information, in the following analyses related to age and gender, we only used the data of users aged between 12 and 40, excluding all users who set their birthdate at Jan. 1, 1990 (which is the default setting of NCM). For this range of ages (12 to 40), in each of the 29 age groups in our sample, we had at least 846 users. In 18 out of 29 investigated age groups our sample had over 5,000 users, and in seven groups over 30,000 individuals, with the maximum number at the age of 21 (a total of 49,870 users).

Given that the year 1990 is set as the default year in NCM apps, we observe in Fig. \ref{track_year}(b) an anomalous peak at the age 26. For the remainig 
user ages, we can see that the number of users increases as they are getting younger, peaking around the age of 21. 

\begin{table}[htbp]
\scriptsize
\centering
 \caption{\label{tab_tagClass} The tag-set of NCM}
 \begin{tabular}{cm{7cm}<{\centering}} 
 \toprule
 Tag Class & Tags \\
\hline
\rowcolor{mygray}
Language & Chinese, EU\&US, Japanese, Korean, Cantonese, Other LNGs \\
Genre & Pop, Rock, Folk, Electronica, Dance, Rap, Light Music, Jazz, Country, R\&B/Soul, Classical, Ethnic, Britpop, Metal, Punk, Blues, Reggae, World Music, Latin, Alternative/Indie, New Age, Antique, Post-Rock, Bossa Nova\\
\rowcolor{mygray}
Scenario & Early Morning, Night, Studying, Working, Noon Recess, Afternoon Tea, Metro, Driving, Sports, Travel, Walking, Bar\\
Emotion & Nostalgia, Refreshing, Romantic, Sexy, Sad, Healing, Relaxing, Lonely, Touched, Exciting, Happy, Quiet, Missing \\
\rowcolor{mygray}
Theme & OST, ACG, Campus, Game, 1970s, 1980s, 1990s, Web Song, KTV, Classic, Cover, Guitar, Piano, Instrumental, Children, Ranklist, 2000s\\
  \bottomrule
 \end{tabular}
\end{table}

As shown in the inset of Fig. \ref{track_year}(b), the sex ratio of users varies by age, with the ratio of male-to-female users increasing significantly with age. This ratio increases faster among users who have exceeded 25 years of age. The length of the FPs -- the number of favorite songs a user saves -- also changes with age. In the group of users below the age of 25, the playlist length is similar among male and female listeners, whereas among the users older than 25, males have significantly longer lists of songs than their female counterparts. In other words, NCM seems to attract more younger female users and more middle-aged male users.

Users can assign different tags to each General Playlist they created, with no more than three tags selected from a given tag-set of NCM. The tag-set includes five classes: Language, genre, scenario, emotion, and theme. Each class further includes several tags, as shown in Table \ref{tab_tagClass}. Favorite Playlists cannot be assigned any tags.

\subsection{User-music bipartite networks}

\begin{figure}
\begin{center}
\includegraphics[width=7.5cm]{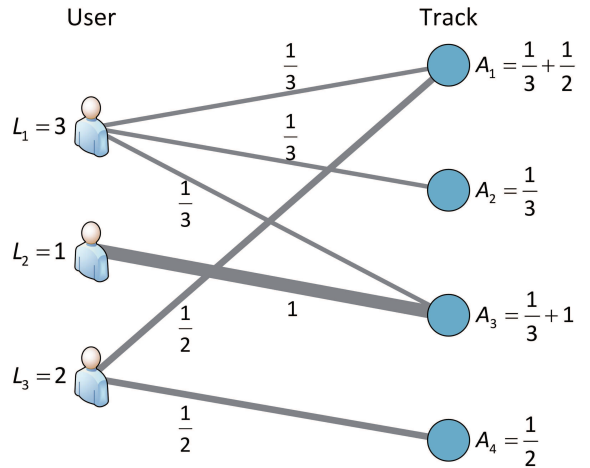}
\end{center}
\vspace{-14pt}
\caption{(Color online) Schematic representation of the user-music bipartite network and the calculation of the total attention $A$ given to each song. The thickness of each gray line is proportional to the ratio of the user's attention focused on a given song.
}\label{bi-network}
\end{figure}

Since each FP contains the favorite songs of a user, these FPs allow us to construct and measure bipartite networks between 
users and songs \cite{gufandi,Zhu2015}. In the bipartite network, each user corresponds to an FP and connects to every track in the generated FP, as shown in Figure \ref{bi-network}.

A bipartite network consists of nodes that can be divided into two disjoint and independent node sets, such that every network link connects from one node set to the other (see Fig. \ref{bi-network}). For example, a user-content bipartite network model has been employed recently to measure users' influence and its diffusion in online social communities such as Pinterest \cite{Zhu2015}. In our case, in the user-music bipartite network of NCM, each user is connected with each track listed in her or his FP, as depicted in Figure 2. We describe the global features of user-music bipartite networks based on three parameters: 1) the length of user's FP corresponding to how many songs a user saves in the FP, 2) the number of users for each track, and 3) the total amount of preferences from all users for each respective track.

To estimate user's musical preferences, we assume that each track in a user's FP shares the same preference weight, and each user's preference has equal weight as well. Thus, when considering the total amount of preferences that a user has over L pieces of music in the FP, the amount of the user's preference for each individual track is 1/L. By summing up all users' preferences for a specific track across all different FPs, we can calculate the total amount of users' preferences or the total amount of attention given to one particular track. For example, two users can save the track {\it a} in their FP (see Figure 2). One user has three tracks in her FP, so that track {\it a} receives 1/3 unit of preference from that user; another user has two tracks in the FP, thus the same track receives 1/2 unit of preference from this other user as well. In total, the resulting total preference for the track {\it a}, or the total attention given to it, is $A_1 = 5/6$ (i.e. 1/3+1/2) unit (see the top right track in Fig. 2).

\begin{figure}
\begin{center}
\hspace*{-0.7em}{
\includegraphics[width=8.8cm]{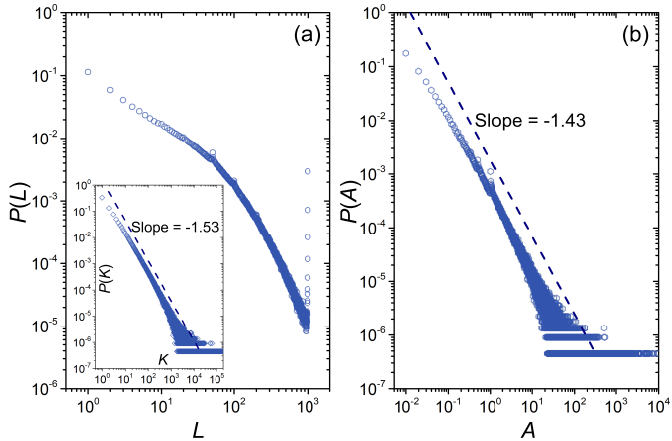}}
\end{center}
\vspace{-14pt}
\caption{(Color online) Three distributions of the user-music bipartite network. (a) The distribution of the length of user's FP. The inset in this figure shows the distribution of the number of followers of each track. (b) The distribution of users' attention focused on each track.
}\label{degree}
\end{figure}

Based on this bipartite network approach and the above calculation of users' attention, the global features of the bipartite network and the patterns of users' attention can be described by the following three distinct distributions: The distribution of the length of a user's FP, the distribution of the number of followers of each track, and the distribution of users' attention $A$ given to each track. Figure~\ref{degree} shows the distribution patterns of these three global features of the user-music bipartite network. First, we see that the patterns of all three distributions are strongly heterogeneous. The distribution of the number of NCM users per song and the distribution of total users' preferences per song, as shown in Figure \ref{degree}, are well fitted by power-law-like functions. These heterogeneous properties have been observed widely as a hallmark of human behavior across many different online social media, indicating that only a few popular songs typically attract the attention of most users, while a relatively small number of active users are actually the carriers of the majority of musical preferences. Interestingly, from the curve damps at $L = 50$, $100$, and $1000$, shown in Figure \ref{degree}(a), we can observe the impact of specific rules set by NCM such as the limits on the maximum length of FPs that were set to 50 and 100 in the previous NCM versions, and to 1000 in the later NCM version.

\subsection{Tag mapping} 

In our analyses related to users' musical preferences, it was necessary to obtain music category information, such as musical genre information. However, besides the tags of General Playlists that are assigned by users, NCM does not provide any further information on a track's category. To obtain the tag information for each track, we map in our analysis the playlist's tags to all the tracks in the playlist, and we proceed with the following tag mapping steps:

i) According to the full tag-set of NCM, to show the initialization of track's tags, we assign to each track in the dataset a vector-set that contains five zero vectors, where each vector corresponds to a tag class, and the number of elements in each vector is equal to the number of tags in its corresponding tag class.

ii) For each track, if a playlist with tag information (e.g. tags A, B, and C) contains a given track, we add a unit value to the track's vector elements that corresponds to tags A, B, and C.

iii) After applying the above steps to all the tracks and all the playlists, we normalize the elements in each non-zero vector in each track's vector-set. For example, if a track's vector of the tag class ``Language'' is $[5, 2, 2, 1, 0, 0]$, its normalized vector is $[0.5, 0.2, 0.2, 0.1, 0, 0]$. Other zero vectors remain unchanged.

A track's tag information can be shown in its corresponding vector-set. For example, for a track with the vector $[0.5, 0.2, 0.2, 0.1, 0, 0]$ of the tag class ``Language'', and with each element in the vector respectively corresponding to the tags ``Chinese, EU \& US, Japanese, Korean, Cantonese, Other Languages'', it would then mean from this particular tag order that the track would be highly related to Chinese music and also slightly related to the music of the EU \& US areas as well as the Japanese music, but a much weaker relationship with Korean or Cantonese music or the music in other languages. In our following discussions, each value in the vector-set is called the tag's strength of the track. Applying this mapping method to all 4,261,266 tracks, we obtained the tag information for a total of 1,464,448 tracks (34.37\%).

\begin{figure}
\begin{center}
\includegraphics[width=8.5cm]{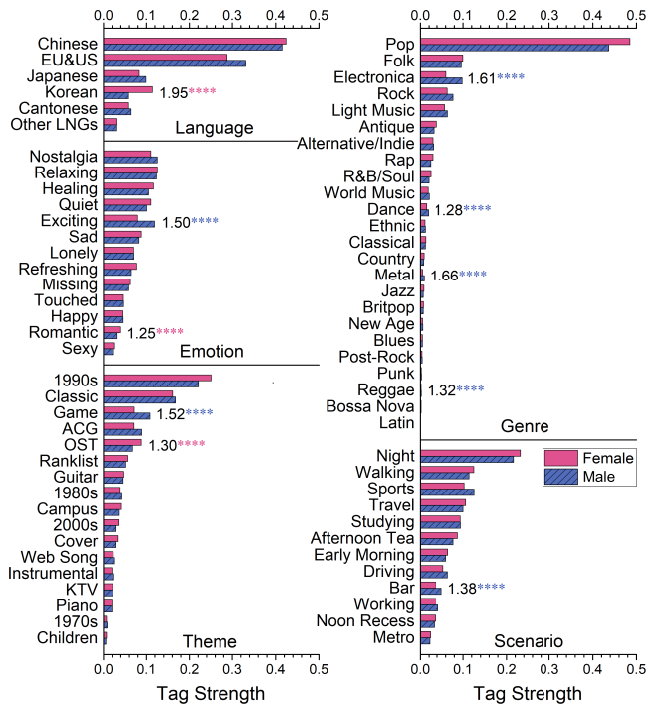}
\end{center}
\vspace{-14pt}
\caption{(Color online) The global tag strength distribution as a function of tag class. The asterisk symbols indicate the tags with large gender difference, with pink asterisks showing the tags for which female users have a higher strength than males, and correspondingly, the blue asterisks indicate the tags with greater strength for male relative to female users. The numbers displayed next to asterisk symbols represent the tag-strength vs. gender ratios.
}\label{tags}
\end{figure}

Similarly, we can further map tracks' tags to the FPs of active users. We also assign each FP a zero vector-set, and add the tag vector-sets of every track in the FP to the FP's vector-set, and we then normalize the elements in each non-zero vector in the FP's vector-set. The FP's vector-set also reflects the user's tag information. This way, we were able to obtain the tag information of 3,021,208 active users (99.76\%).

Furthermore, in our analysis that divided users into distinct groups, we applied a similar tag mapping method that adds up the tag vector-set of each user in the group and normalizes the cumulative vector-set to represent the vector-set of the group. We also obtained the global vector-set that cumulates and normalizes the vector-sets of all users. Figure \ref{tags} shows the tags' strength of the global vector-set for male and female users. Obviously, the distribution of tag strength in the tag classes of ``Genre" is more heterogeneous, with a few genres, such as ``Pop",``Folk " and ``Electronica" attracting most users' preferences. Several tags, as denoted by asterisks in Figure \ref{tags}, are associated with large gender differences. For example, female preferences for the tags ``Korean", ``OST" and ``Romantic", are obviously higher than those of male users, while ``Metal", ``Electronica", ``Game", and ``Exciting" tags show the opposite pattern of gender differences, reflecting the gender divide on musical tastes. In the following analyses, we only focus on FPs, whereas the information on General Playlists is not used anymore.

\subsection{Community detection}

\begin{figure}
\begin{center}
\includegraphics[width=8.5cm]{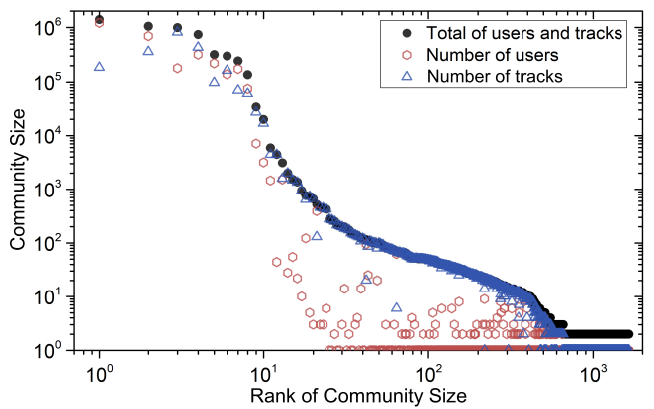}
\end{center}
\vspace{-14pt}
\caption{(Color online) The distribution of the NCM community size in the Zipf's ranking plot. The communities are ranked by the total 
number of music tracks and users. The blue and red data points, respectively, denote the numbers of tracks and users in each community.
}\label{community}
\end{figure}

Although the tag mapping method can reveal a lot about prevalent music genres and corresponding users' preferences, tag mapping 
cannot capture all information on the NCM's tracks and users. Since we have 
established a user-music bipartite network, community analysis, as a network-based clustering-like method \cite{Newman2004}, can serve as an 
efficient approach to exploring tracks' categories and user's musical preferences. However, unlike unipartite networks which have been investigated extensively in past years with a variety of community identification algorithms \cite{berahmand,zwang,kumardohare}, the methods for community detection in bipartite networks have been less frequently studied and applied to real-world networked systems \cite{shihuazhang2016}.

\begin{table*}[htbp]
\scriptsize
\centering
 \caption{\label{tab_commuity} The eight largest communities of NCM. The communities are ranked by size, ranging from large to small. $\rho^T$ and $\rho^U$, respectively, are the proportion of tracks in global total tracks, and the proportion of users in global total users of each community. The rightmost five columns show the communities' first and second primary tags in each of the five tag classes. }
 \begin{tabular}{c@{ } c c  c  c@{ }   c c  c c c} 
 \toprule
 Commuity & $\rho^T$ & $\rho^U$ & Average age & Female proportion & Language &  Genre & Scenario & Emotion & Theme\\
\hline
\rowcolor{mygray}
A & 0.0819 & 0.4006 & 22.7 & 0.337 & Chinese, Cantonese  & Pop, Folk & Night, Working & Nostalgia, Sad & 1990s, Classic\\
B & 0.1582 & 0.2297 & 21.7 & 0.289 & EU\&US, Chinese & Electronica, Pop & Sports, Night & Exciting, Relaxing & Game, 1990s\\
\rowcolor{mygray}
C & 0.3615 & 0.0591 & 24.8 & 0.299 & EU\&US, Chinese & Light Music, Rock & Night, Studying & Quiet, Relaxing & OST, Classic\\
D & 0.1903 & 0.1048 & 20.5 & 0.331 & Japanese, Chinese & Light Music, Antique & Night, Studying & Healing, Relaxing & ACG, Game\\
\rowcolor{mygray}
E & 0.0430 & 0.0729 & 20.7 & 0.667 & Korean, EU\&US & Pop, Rap & Night, Sports & Relaxing, Healing & 1990s, OST\\
F & 0.0719 & 0.0458 & 28.7 & 0.247 & Chinese, EU\&US & Pop, Ethnic & Night, Driving & Nostalgia, Quiet & Classic, OST\\
\rowcolor{mygray}
G & 0.0314 & 0.0569 & 22.0 & 0.352 & Chinese, EU\&US & Folk, Rock & Night, Travel & Quiet, Lonely & Guitar, Classic\\
H & 0.0271 & 0.0249 & 23.0 & 0.423 & EU\&US, Other LNGs & Pop, Ethnic & Driving, Night & Relaxing, Exciting & Classic, 1990s\\
  \bottomrule
 \end{tabular}
\end{table*}

In a typical complex network, some nodes can be identified as a group if they are densely connected with each other while their connections outside of the group are relatively sparse. This type of node grouping in a network is called a ``community", and we say that groups of nodes form a community structure within the network. Generally speaking, a community structure can be identified if network nodes can be grouped into sets such that each set of nodes 
exhibits dense internal connections \cite{Newman2004}. For example, users with similar musical tastes are more likely to save songs from 
same or similar genres in their FPs; thus, songs from the same or related genres would typically appear within the same community of the user-music bipartite network. Therefore, by investigating the community structure of songs in the bipartite network, we can measure the diversity of musical genres that are typically preferred by the NCM users. It is important to note here that the identification using community structure is completely socially-determined, because even though two types of music may have somewhat different style characteristics they would still be identified as belonging to the same community if they share similar listeners.

We detect the listenership communities in our user-music NCM network by employing the fast-unfolding algorithm introduced by Blondel et al. \cite{Blondel2008}. This algorithm retains both the higher accuracy and the lower algorithmic complexity, and is widely used in network-based analyses 
of social media \cite{Lambiotte2009}. Without differentiating each side of the nodes (either music or users) in this bipartite network, the fast-unfolding algorithm would identify a community that includes at least one user and one song that was included in the user's FP. In total, there 
were 1656 initially detected communities in this user-music bipartite network. As shown in Figure \ref{community}, the size distribution of these detected communities is highly heterogeneous, such that 99.47\% of users and 96.53\% of tracks can be grouped into eight largest communities. Our Table \ref{tab_commuity} shows the information related to these eight largest communities. 

In addition, using the above tag mapping method, we calculated the tag vector-set of the identified eight largest communities. For each community, the tag with the largest strength in each tag class is called the first primary tag, and the second largest one is the second primary tag. As shown in Table \ref{tab_commuity}, the first and the second primary tags of each community are very different, possibly reflecting the individual differences in musical preferences, the underlying moods, or even the wellbeing status of users in different communities.

\subsection{Diversity analysis}

Diversity analysis has been used widely in the study of musical preferences \cite{Park2016}. Following Eagle et al. \cite{Eagle2010}, and based on the tag information and the communities detected in our dataset, we define a series of metrics to describe and assess various types of diversity of musical preference behaviors.
For the $i$-th user, given her normalized tag vector $(\upsilon_{i1}, \upsilon_{i2},...)$ of a given tag class (e.g. tag class $G$), her individual tag diversity $D^{TI}_{Gi}$ of the tag class $G$ is then defined as \cite{Wangetal2016}:
\begin{equation}
D^{TI}_{Gi} = \frac{-\sum_j\upsilon_{ij}\log(\upsilon_{ij})}{\log{m_G}},
\end{equation}
where $\upsilon_{ij}$ is the tag strength of the $j$-th tag of tag class $G$, and $m_G$ is the total number of tags in the tag class $G$; for example, $m_G = 6$ for the tag class ``Language". The numerator of the expression is actually the information entropy of the distribution of each element in the tag vector, and this diversity measure addresses the degree of homogeneity of a user's preference for a given music tag in a given tag class. For example, in the tag class $G$, if the user listens only to the music with a single tag, then there is only one tag with a non-zero strength in the user's tag vector of the tag class $G$, and $D^{TI}_{Gi}$ equals the minimum value of $0$, indicating that the user's preference for the tag class is quite simple; moreover, if each tag has the same tag strength, $D^{TI}_{Gi}$ equals the maximum value $1$, indicating that the user has equal preferences for the music of each tag in the tag class.
For a given demographic group, the average value of $\langle D^{TI}_G\rangle$ of tag class $G$ describes the average level of the tag diversity across all users' individual preferences in that group.

Similarly, for a given demographic group of users, obtaining the group's normalized tag vector $(g_{G1}, g_{G2},...)$ of the tag class $G$, the aggregated tag diversity $D^{TA}_G$ of the tag class $G$ is defined as:
\begin{equation}
D^{TA}_G = \frac{-\sum_jg_{Gj}\log(g_{Gj})}{\log{m_G}}, 
\end{equation}
where $g_{Gj}$ is the tag strength of the $j$-th tag in the tag class $G$ of the group's tag vectors. It describes the tag diversity of the group's users in an aggregated fashion, namely, $D^{TA}$ treats users in a certain demographic group as a whole.

In order to analyze the differences among users within the same demographic group, we need to understand the relationship between individual diversity $D^{TI}$ and the aggregate diversity $D^{TA}$. If all users in a certain demographic group like tracks with similar tags, the value of $D^{TA}$ will be close to $D^{TI}$. If all users' preferences for tags in the same group are different, the gap between the $D^{TI}$ and $D^{TA}$ would be large. Therefore, the aggregated tag diversity $D^{TA}$ of a certain demographic group contains two components: The individual tag diversity of each user, and the within-group diversity of users. We thus define the within-group diversity of users for a tag class $G$ as:
\begin{equation}
D^{TU}_G = D^{TA}_G - \langle D^{TI}_G\rangle.
\end{equation}
It measures the difference among users within the same demographic group on a given tag class.

Similarly, we can define three types of community diversities based upon the detected communities in the investigated NCM network.
For the $i$-th user, her individual community diversity $D_i^{CI}$ is given by:
\begin{equation}
D^{CI}_{i} = \frac{-\sum_jv_{ij}\log(v_{ij})}{\log{M_c}},
\end{equation}
where $v_{ij}$ is the proportion of the $j$-th community's tracks in the $i$-th user's FP, and $M_c$ is the global, total number of communities. $D^{CI}_{i}$ equals to $0$ if the user only follows the tracks within the same community, and $D^{CI}_{i}$ equals to $1$ if the user's FP tracks 
distribute homogeneously across each of the existing communities. Thus, unlike individual tag diversity, individual community diversity measures the degree of homogeneity with which the user's FP tracks distribute across different communities. The higher the value, the more homogeneous is the distribution of FP tracks across communities; correspondingly, the lower the value, the more concentrated are the user's FP tracks within only a few communities, representing thus the less diversified preference. The value of $\langle D^{CI}\rangle$ describes the community diversity when averaged across all users' individual preferences in the group.

For a certain demographic group, we further define the aggregated community diversity as:
\begin{equation}
D^{CA}= \frac{-\sum_jq_{j}\log(q_{j})}{\log{M_c}}, 
\end{equation}
where $q_{j}$ is the proportion of the $j$-th community's tracks in all FPs of the users in the group: 
\begin{equation}
q_{j} = \frac{\sum_kl_{kj}}{\sum_k L_k}, 
\end{equation}
where $l_{kj}$ is the total number of the $j$-th community's tracks in the $k$-th user's FP, $L_k$ is the total number of tracks in the $k$-th user's FP, and $\sum_k$ denotes the summation over all users in the group. This metric describes the aggregated diversity of users' preferences for tracks in a certain demographic group. A higher value indicates a more diversified taste in music of a certain group of users.

Similar to the case of tag diversities, the aggregated community diversity $D^{CA}$ of a certain demographic group also contains the individual tag diversity of each user and the within-group diversity of users. We therefore use the difference:
\begin{equation}
D^{CU} = D^{CA} - \langle D^{CI}\rangle,
\end{equation}
to measure the within-group community diversity of users.

\subsection{Group Difference Analysis}

The Kullback-Leibler divergence (KLD) is an extensively employed measure for a difference between two probability distributions \cite{Kullback1951}. Generally, the KLD from a normalized discrete distribution ($P$) to the reference normalized discrete distribution ($Q$) is: $\Delta(P||Q) = \sum_i P(i) \ln \frac{P(i)}{Q(i)}$ for each $i$ satisfying $P(i) > 0$ and $Q(i) >0$. The value of KLD is non-negative. A KLD of 0 indicates that the two normalized distributions are exactly the same. A higher KLD means a larger difference between the two distributions. Its value of 1 means that the two 
distributions are so different that we cannot approach one of them from the other. Since $\Delta(P||Q)$ generally does not equal $\Delta(Q||P)$, the form employed in our analysis is the mean value of $\Delta(P||Q)$ and $\Delta(Q||P)$:
\begin{equation}
\begin{aligned}
\Delta &= \frac{1}{2}\left[\Delta(P||Q) + \Delta(Q||P)\right] \\
       &= \frac{1}{2}\left[\sum_i P(i) \ln \frac{P(i)}{Q(i)} + \sum_i Q(i) \ln \frac{Q(i)}{P(i)} \right].
\end{aligned}
\end{equation}

Our group difference analysis mainly focuses on gender differences in users' musical preferences. Similar to the case of the diversity analysis, gender differences have two types of representations: One is the difference between the vector-sets of tags of the two groups, and the other is the difference between the community distributions of users' FP tracks in the two groups. The KLD, therefore, has two different definition levels in our analysis: One is tag-based and the other one is based upon the detected communities in our investigated NCM dataset.

For each studied demographic group, we separate male and female users into two subgroups. For the male user subgroup with its normalized tag vector $(g^M_{1}, g^M_{2},...)$ of the tag class $G$, and for the female user subgroup with its normalized tag vector $(g^F_{f1}, g^F_{2},...)$ of the tag class $G$, the associated tag-based KLD measure for gender differences within the tag class $G$ is defined as:
\begin{equation}
\Delta^T_G = \frac{1}{2}\left[\sum_i g^M_{i} \ln \frac{g^M_{i}}{g^F_i} + \sum_i g^F_i \ln \frac{g^F_i}{g^M_i}\right].
\end{equation}

Similarly, for a given demographic group, obtaining the normalized proportion $q^M_{i}$ of the $i$-th community's tracks in all users' FPs in the male subgroup, and the corresponding normalized proportion $q^F_{i}$ for the subgroup of female users, the community-based KLD for gender differences can be calculated by:
\begin{equation}
\Delta^C = \frac{1}{2}\left[\sum_i q^M_i \ln \frac{q^M_i}{q^F_i} + \sum_i q^F_i \ln \frac{q^F_i}{q^M_i}\right].
\end{equation}

\section{Results and discussions}

\subsection{The decay of music influence}

Most cultural products are typically characterized by their life cycle and the associated diffusion pattern related to their ability to attract people's attention. For example, the influence of a novel cultural product usually attains its maximum peak after a diffusion process, and then trends to a long-term decay. We expected that the evolution of music popularity would show similar features: Given the potentials of modern online social media with rapid information dissemination, the term of the diffusion process of a novel musical product before the peak should generally be shorter than its subsequent decay stage. With this consideration in mind, our analysis of the evolution of music popularity focuses on the mode of the decay term of music influence.

\begin{figure}[h]
\begin{center}
\includegraphics[width=8.5cm]{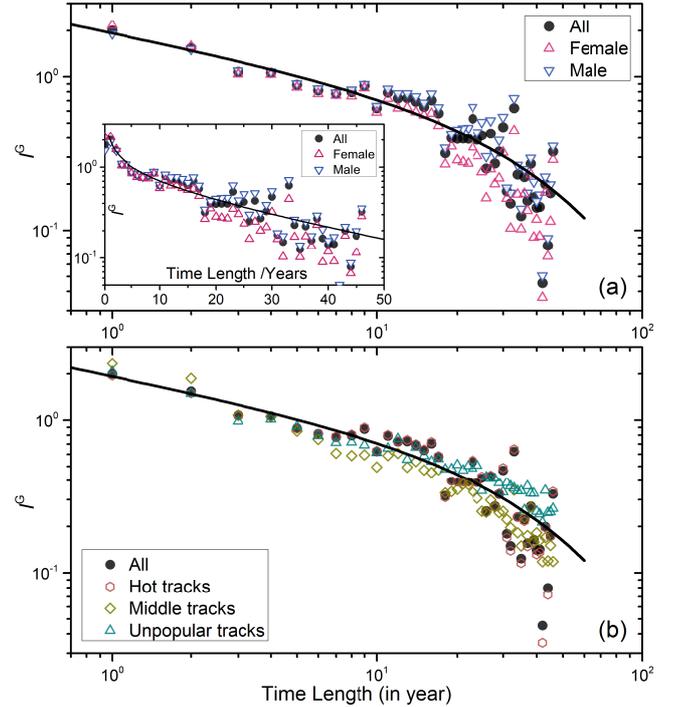}
\end{center}
\vspace{-14pt}
\caption{(Color online) The variation of the average global preference $I^G$ of tracks as a function of the time interval between the release year of the music track and the data collection year (2016) in a log-log plot. Panel (a) shows the associated gender differences; the inset depicts the same data in a semi-log plot. Panel (b) shows the decay mode for the tracks with different attention levels. The solid line in each panel represents the fitting 
function for all tracks, which is power-law-like with the exponential tail $y = 1.97x^{-0.34}\exp(-0.023x)$. 
All tracks published in the same year were ranked in accordance with their total attention received; the top 5\% of tracks in this attention ranking are the "hot tracks", the tracks ranked between the top 5\% and top 25\% are the "middle tracks", and the rest are the "unpopular tracks". 
}\label{decay}
\end{figure}

In the NCM dataset, each track belongs to an album, and the dataset includes the release year of each album. The dataset also includes users' birth dates. Similar to the method used to calculate a track's total attention $A$, we now define a matrix $\bm{A}^Y$ to describe the attention map between users with different birth years and tracks with different release years.

Each element of the matrix $\bm{A}^Y$ is then:
\begin{equation}
A^Y_{ij} = \frac{1}{n_j}\sum_N\sum_{L_k}\frac{\varepsilon_{ij}}{L_k}
\label{matrixA}
\end{equation}
where $i$ and $j$ denote the $i$-th release year of a track and the $j$-th birth year of a user, $N$ is the total number of users in the dataset, $L_k$ is the number of tracks in the $k$-th user's FP, and $n_j$ is the total number of users born in the $j$-th year. $\varepsilon_{ij} = 1$ if the track in the FP of the $k$-th user was released in the $i$-th year and the $k$-th user was born in the $j$-th year; otherwise $\varepsilon_{ij} = 0$. Obviously, each column of $\bm{A}^Y$  is actually a normalized distribution of the release year of the favorite tracks for the users with the same birth year.

Similarly, the global distribution of the release years of favorite tracks of all users can be obtained as:
\begin{equation}
A^G_{i} = \frac{1}{N}\sum_N\sum_{L_k}\frac{\varepsilon_{i}}{L_k},
\end{equation}
where $N$ is the total number of users, and $\varepsilon_i = 1$ if the track in the FP of the $k$-th user is released in the $i$-th year, and otherwise $\varepsilon_i = 0$. The global distribution describes the average level of users' preferences for the tracks that were released in different years.

\begin{figure}
\begin{center}
\includegraphics[width=8.6cm]{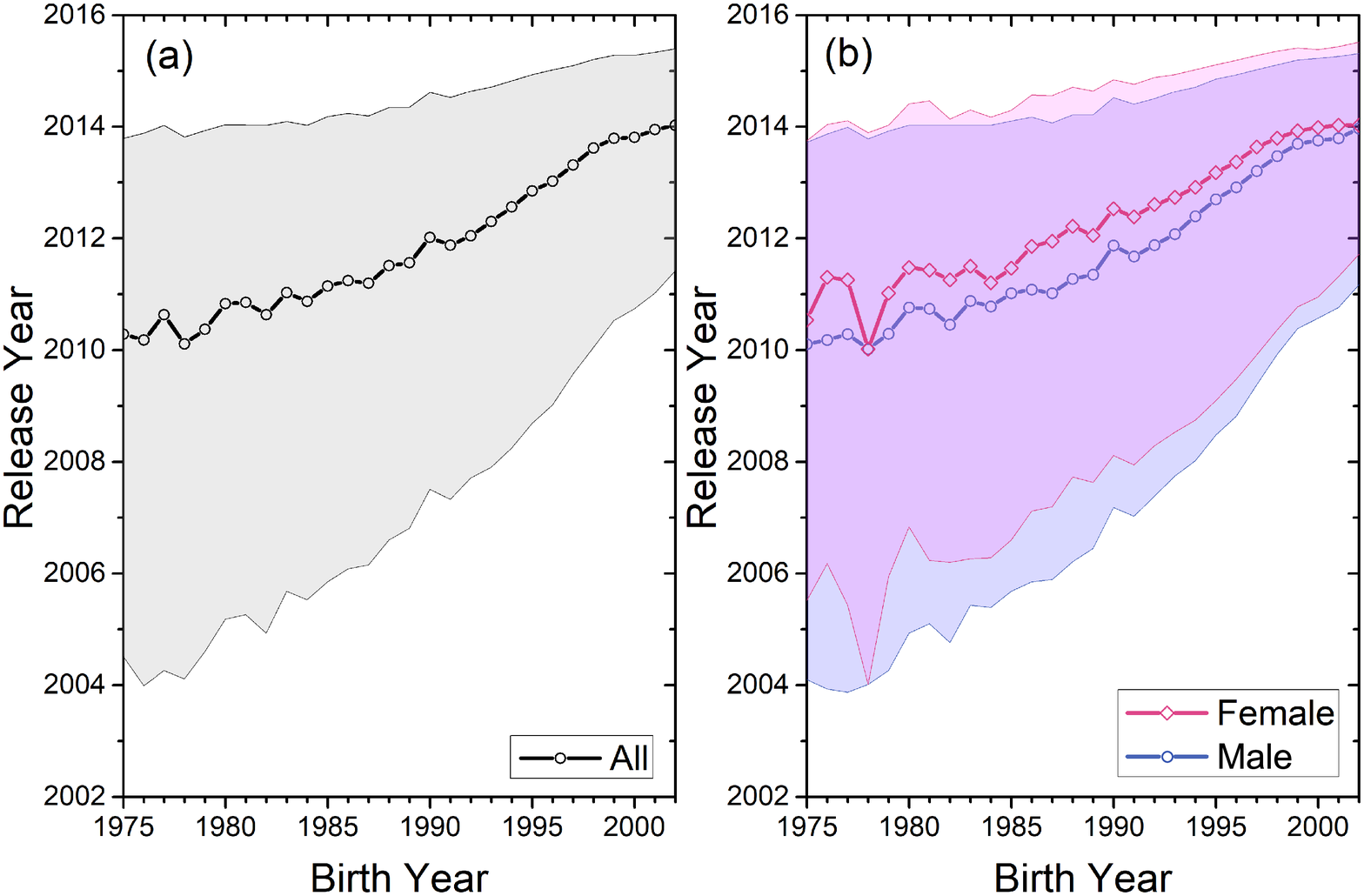}
\end{center}
\vspace{-14pt}
\caption{(Color online) The release year distribution of user's FP tracks for the users with a different birth year. Panel (a) shows the global pattern, where the dotted-line stands for the median of the distribution. Panel (b) shows the corresponding patterns of female and male users, and the dotted-lines depict the medians of each distribution. The shadowed areas (light blue shadow for male users, light pink shadow for female users, and the light purple shadow for their overlap) show the regions between the corresponding upper and lower quartiles.
}\label{median}
\end{figure}

Since the total number of tracks released in different years can vary, we calculate the average global preference as:
\begin{equation}
I^G_{i} = \frac{A^G_i}{\rho_i}=\frac{1}{\rho_i N}\sum_N\sum_{L_k}\frac{\varepsilon_{i}}{L_k},
\end{equation}
where $\rho_i$ is the ratio of the number of tracks released in the $i$-th year, namely, $\rho_i = m_i/M$, where $m_i$ is the total number of tracks released in the $i$-th year, and $M$ is the total number of tracks. Generally, the average preference for old tracks is much lower than for recently released tracks. Naturally, older users show higher preferences for older tracks.

We next addressed the relationship between users' age and the release year of their preferred music tracks, as well as the question of how fast does a user's taste in music decay over time. In Fig. \ref{decay}, we re-plot the curve of $I^G$ by rescaling the x-axis to the difference between 2016 and the 
track release year. Fig. \ref{decay} thus shows the decay of users' musical preferences: The pattern is well fitted by a power-law-like function with the exponent tail $y = 1.97x^{-0.34}\exp(-0.023x)$. Such power-law-like decaying patterns for musical preferences did not reveal a significant difference between male and female users (Fig. \ref{decay}(a)), nor did they show significant differences among tracks with varying popularity (Fig. \ref{decay}(b)).

Remarkably, similar slow-falling patterns with a power-law-like tail have recently been identified in the temporal variation of different 
online behaviors, e.g. in the time series characterizing the outbreak of internet memes or in the spreading patterns of popular topics \cite{Crane2008,Leskovec2009}. However, a noticeable difference 
is that the decay of music preferences that we observed in the NCM dataset is much slower than in any other case previously observed in online social media, which is in part corroborated by a more recent study on the online attention given to cultural products such as movies, biographies, 
and popular songs \cite{Candia2019}. 

The value of the characteristic exponent of the power-law function for musical preferences in our study is 0.34, which is also lower than what was typically reported for other cases in online social media. The exponent tail denoting the stage of rapid decay often emerges when the time period is longer than 20 years, which closely corresponds to one generation, thus implying that the rapidly decaying cultural taste is a generational phenomenon. Within the same generation, the cultural preferences are much more stable.

\subsection{The relationship between users' age and sensitivity to music}

Age and gender have often been identified as important factors in shaping musical attitudes and preferences \cite{Bonneville2013,Bonneville2017a}. 
A common support for this influence of demographic factors on preference behavior is e.g. the observation that teenagers typically 
pay far more attention than adults to new pop music trends \cite{Bonneville2013}.

To further analyze age trends in the context of Chinese users' musical preferences, we calculated the median, 25\%, and 75\% quartiles of their 
preferences over the songs that were released across different years spanning the period of several decades. Figure \ref{median}(a) shows the central region of the area between the upper and lower quartiles of preferences by the age cohort. For older listeners, we can see that they clearly prefer older songs covering a wider range of release years than younger listeners. On the other hand, younger listeners evidently pay more attention to more recently released tracks. Figure \ref{median}(b) depicts the associated gender differences. Even though they both display a similar trend, especially around and before the age of 20, women are more likely to prefer more popular and newly-released songs than men, suggesting that female users are more sensitive than men to the influence of musical trends. This finding is in line both with earlier offline studies of musical preferences \cite{Rawlings1997} as well as with the more recent analyses of large-scale music-based online communities 
such as Spotify \cite{Kalia2015}. To control for cross-cultural effects, the study in Ref. \cite{Kalia2015} investigated only US Spotify users. 
Excitingly, this remarkable similarity between our findings for NCM Chinese music listeners and the results for US Spotify users suggests the existence of cross-cultural universals in preferences for popular music that hold across a wide variety of demographic factors.

\begin{figure*}[htb]
\begin{center}
\hspace*{-0.7em}
\includegraphics[width=15.3cm]{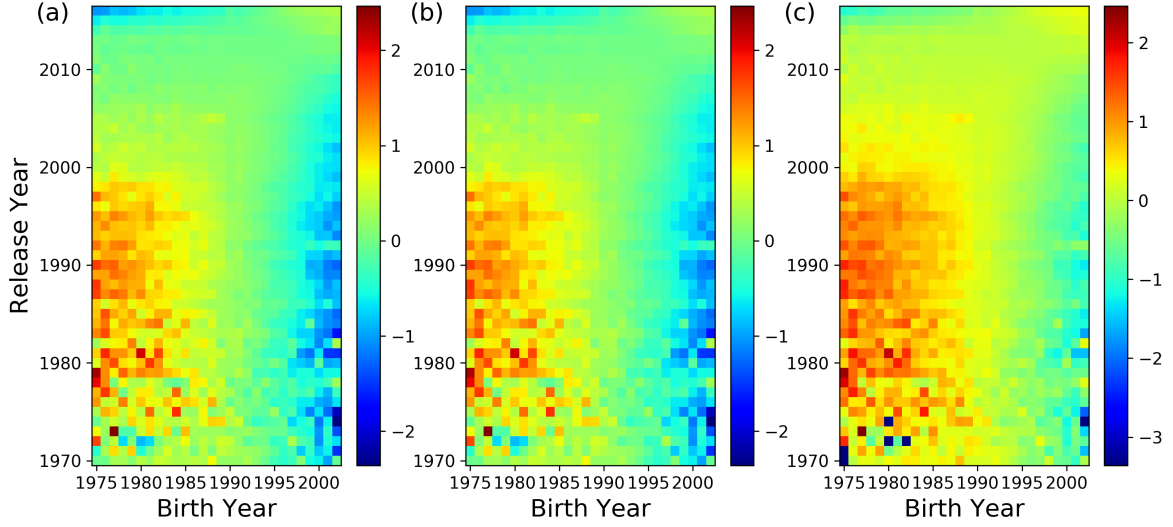}
\end{center}
\vspace{-14pt}
\caption{(Color online) The heat maps of the matrix $\ln(\bm{R})$: (a) the global pattern, (b) the pattern of male users, and (c) the pattern of female users.
}\label{matrixR}
\end{figure*}

As a further part of our analysis, we investigated the musical sensitivity of NCM users. We defined the matrix $\bm{A^Y}$ shown in Eq. (\ref{matrixA}) to describe the distribution of users' attention to the tracks released in year $i$ for the users with birth year $j$, and the global attention
distribution $A_i^G$. Here, we treat $A^G$ as the background distribution of user's attention, and we use the matrix $\bm{R}$ of the ratio between $A^Y_{ij}$ and $A_i^G$ to express the relative attention of users with a birth year $j$:
\begin{equation}
R_{ij} = \frac{A^Y_{ij}}{A^G_i},
\end{equation}
representing thus the attention of users relative to the global average level.
The maps of the matrix $\bm{R}$ for all users and the two corresponding cases for male and female users are shown in Figure \ref{matrixR}.
We see again that users' preferences generally depend upon their age: Older users pay considerably more attention to older tracks.

For the case $i > j$, the value $i - j$ actually denotes the age of users when the tracks were released. We therefore define the sensitivity $S_{j}(i - j)$ of users with birth year $j$ for the tracks released in year $i$, as the normalized value of $R$ for each $j$:
\begin{equation}
S_{j}(i-j) = R_{ij}\langle R_{kj}\rangle^{-1}=\frac{A^Y_{ij}}{A^G_i} \left\langle \frac{A^Y_{kj}}{A^G_k}\right\rangle^{-1},
\end{equation}
where $k$ denotes all the release years of tracks, and angle brackets represent the operator for the average value. Here we set that $S = 1$, corresponding to the case where the users' attention is equal to the global average value of attention.

\begin{figure}
\begin{center}
\hspace*{-0.9em}
\includegraphics[width=8.2cm]{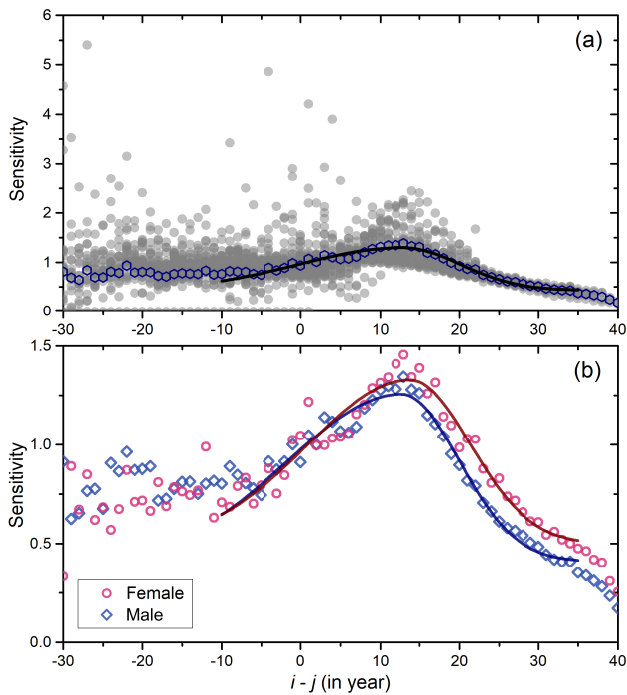}
\end{center}
\vspace{-14pt}
\caption{(Color online) (a) The sensitivity of NCM users to the tracks released at their different ages, whereby negative age represents the cases in which tracks were released before the birth of the user. The blue open circles are the average values of all data points (gray dots) of sensitivity, and the black line shows the fitting Bigaussian function. (b) Gender difference for the averaged sensitivity data as a function of users' age when the tracks were released. The dark blue and dark red lines respectively are the fitting Bigaussian functions for the cases of male and female users.
}\label{sensitivity}
\end{figure}

Figure \ref{sensitivity}(a) shows the users' sensitivity to music as a function of users' age when the tracks were released, 
whereby negative numbers correspond to the track releases before the users' birth year. The average sensitivity of users of each age is depicted 
by the open blue circles which exhibit a unimodal shape that can be nicely approximated by a Bigaussian function:
\begin{equation}
y(y_0,x_c,H,w_1,w_2) =
 \begin{cases}
    y_0 + H\exp{[-0.5(\frac{x-x_c}{w_1})^2]},   &  \text{$x < x_c$} \\
    y_0 + H\exp{[-0.5(\frac{x-x_c}{w_2})^2]},
        &  \text{$x \geq x_c$}
 \end{cases}
\end{equation}

\noindent where the estimated values of the parameters $y_0$, $x_c$, $H$, $w_1$ and $w_2$, respectively, are 0.43, 12.88, 0.87,13.18 and 7.26. 
The peak of the fitting curve thus corresponds to the value $x_c = 12.88$, meaning that on average, the user's sensitivity to music reaches its peak around the age of 13. Thus, people's sensitivity to music is the greatest when they are in their teen ages. This finding resonates well with the common experience that people tend to be most 
strongly influenced by the music that has been popular in their youth. Moreover, this finding is corroborated by a recent big data analysis of US 
Spotify users \cite{Stephens-Davidowitz2018}, reporting the strongest influence on adult musical preferences at the ages of 13 (for women) and 14 (for men). 

This striking similarity between our findings and those observed with US Spotify listeners highlights again the universal nature of musical preferences that holds across cultures, this time with respect to the period of the formation of adult musical tastes. We further observed that the fitting function is not symmetrical, and the decay of sensitivity after $x_c$ occurs 
faster than its growth before $x_c$ ($w_1 > w_2$), indicating a rapid decline in sensitivity to new music after adolescence.

We further observed an interesting relationship between gender and users' musical sensitivity. As shown in Figure \ref{sensitivity}(b), 
the fitting curves for the average sensitivity of men and women are different. The estimated values of the parameters $y_0$, $x_c$, $H$, $w_1$ and $w_2$, respectively, are 0.41, 12.76, 0.85, 14.25 and 7.19 for male users, and 0.50, 13.79, 0.83, 12.83 and 7.60 for female users. 

On average, men thus reached their peak of sensitivity to music at a younger age than women; however, the estimated peak age $x_c$ of male users was only slightly smaller than that of their female counterparts. Moreover, after passing the peak-age of sensitivity to music, women remained more sensitive than men at the same age to the most recent music trends. These results are further in line with what is revealed 
in Figure \ref{median}(b), namely, that adult female users are usually more sensitive to newly released music, while male listeners develop an earlier and stronger 'taste freeze' phenomenon. 

Together, these findings are partly corroborated by the study of US Spotify users \cite{Kalia2015}, showing that men's preferences for popular music decrease faster than those of their female counterparts, lending thus support for the existence of a further cross-cultural universality in music taste formation. However, a unique, culture-specific Chinese feature in the investigated data could be the observed earlier peak for men than for women, 
since US music listeners showed either roughly equal patterns for both male and female listeners in their teens \cite{Kalia2015}, or the reverse pattern, with women showing an earlier music sensitivity peak than men \cite{Stephens-Davidowitz2018}. However, this uniqueness of Chinese listeners in our analysis may also 
be due to limitations in the available dataset, such as the generally larger male subpopulation that was present in our studied sample of NCM users.

\subsection{Genre-related age trends and gender differences}

Besides discovered general age trends for popular music, age and gender as demographic factors may also exhibit genre-specific influences on musical preference behavior. To explore the genre-related age trends in users' musical preferences, we firstly calculated the tag-vectors 
of each age group, and we then divided the age ranges from 12 to 40 into three separate age stages: 12-18, 19-25, and 26-40, and finally, we averaged the tag strength within each age stage for each tag. Furthermore, for each tag, we compared the average tag strengths for the pairs of two continuous age stages. If the latter is higher than the former 10\%, we used the symbol ``$\verb|/|$'' to express the change trend between the two age stages. We further used ``$\verb|\|$'' if the latter is lower than the former 10\%, and ``$-$" for the rest of the cases. This way, we can obtain the age modes of each tag, which is expressed by combining two symbols from ``$\verb|/|$'', ``$-$" and ``$\verb|\|$''. For example, ``$\verb|//|$'' means that the strength of the tag continuously increases with age, and ``$\verb|/\|$'' denotes a bell-shape-like trend for age, whereas ``$--$'' means that the tag's strength remains largely unchanged across the three stages.

\begin{table*}
\scriptsize
\hspace*{-2.4em}
 \caption{\label{tab_Agemode} The mode of age trends of each tag for male and female users}
 \begin{tabular}{@{} >{\columncolor{mywhite}}c c@{ } m{1.9cm}<{\centering}@{ } m{2.2cm}<{\centering}@{ } m{1.7cm}<{\centering} m{3.2cm}<{\centering}@{ } m{1.0cm}<{\centering}@{ } m{2cm}<{\centering}@{ }m{1.25cm}<{\centering}@{ } m{1.25cm}<{\centering}} 
\toprule
\multicolumn{1}{>{\columncolor{mywhite}}l}{Gender} & Tag Class & $\verb|//|$ & $\verb|/|- $ & $- \verb|/|$  & $- -$ &  $\verb|\/|$ & $\verb|\|-$ & $- \verb|\|$ & $\verb|\\|$ \\
\hline
\rowcolor{mygray}
\multicolumn{1}{>{\columncolor{mywhite}}l}{} & Language & Other LNGs  & Cantonese & Chinese  & EU\&US  &   &   &   & Japanese, Korean\\
& Genre &  Ethnic, Jazz, Blues, Punk, Latin  & Folk, Rock, Country, Britpop, Post-Rock, Bossa Nova & World Music, Classical, Metal, New Age & Pop, Light Music, Alternative/Indie &  & Antique, Dance, Reggae & R\&B/Soul & Electronica, Rap\\
\rowcolor{mygray}
\multicolumn{1}{>{\columncolor{mywhite}}l}{Female} & Scenario &  & Travel & Driving, Working  & Night,Walking, Afternoon Tea, Early Morning, Bar, Noon Recess, Metro &  & Sports, Studying &  & \\
& Emotion & Nostalgia & Lonely, Missing &  & Relaxing, Healing, Quiet, Sad, Touched, Romantic, Sexy & & Refreshing, Happy &  & Exciting\\
\rowcolor{mygray}
\multicolumn{1}{>{\columncolor{mywhite}}l}{}& Theme & Classic, 1980s, Instrumental, 1970s & Guitar, KTV & Children  & OST, Campus, Piano  & Web Song & Cover & 1990s, Ranklist & Game, ACG, 2000s\\
\hline
&Language & Chinese & Cantonese  & Other LNGs  & &  & EU\&US & Korean & Japanese \\
\rowcolor{mygray}
\multicolumn{1}{>{\columncolor{mywhite}}l}{}& Genre & Jazz, Blues, Latin & Folk, Country & Rock, Ethnic, Classical, New Age, Punk   & Pop, Antique, R\&B/Soul, World Music, Britpop, Post-Rock, Bossa Nova  & Metal  & Light Music, Alternative/Indie, Dance, Reggae &  & Electronica, Rap\\
\multicolumn{1}{>{\columncolor{mywhite}}l}{Male} & Scenario &  & Night, Travel, Afternoon Tea, Metro  &    & Walking, Early Morning, Driving, Working, Noon Recess &   & Studying, Bar &  & Sports \\
\rowcolor{mygray}
\multicolumn{1}{>{\columncolor{mywhite}}l}{}& Emotion & Nostalgia & Quiet, Sad, Lonely, Missing &   & Relaxing, Healing,  Refreshing, Touched, Romantic, Sexy &  & Happy &  & Exciting\\
&Theme & Classic, Guitar, 1980s, 1970s & Campus, KTV & Instrumental, OST, Children  & 1990s, Ranklist, Cover, Piano  &   & Web Song &   & Game, ACG, 2000s\\
  \bottomrule
 \end{tabular}
 \begin{tablenotes}
    \footnotesize
    \item None of the tags were observed to show the trend ``$\verb|/\|$'', which was thus not included in this table. The full description of trends is given in Section 3.3.
\end{tablenotes}
\end{table*}

The age modes of tags for each tag class are listed in Table \ref{tab_Agemode}. Generally, age modes with upward trends ( e.g. $\verb|//|$, $\verb|/|- $ and $- \verb|/|$) indicate that a tag is more likely to attract older users, whereas downward-trend age modes (e.g. $\verb|\|-$, $- \verb|\|$, and $\verb|\\|$) mean that a tag would attract more younger users. These age modes of tags reveal rich information about musical age trends, cultural influences, users' situational contexts, and pervading emotions in different age groups. There are several examples: Some new-rising genres or themes, such 
as ``Electronica'', ``Game'', or ``ACG'' have a decaying age mode, suggesting that these music types mainly attract young listeners. In contrast, traditional 
or sophisticated music genres (e.g. ``Classical'', ``Ethnic'' and ``Jazz'') mainly show growing age modes, which is generally in agreement with findings reported in Ref. \cite{Bonneville2017a}. The decaying age modes of tags ``Japanese'' and ``Korean'' indicate the rapidly-growing cultural influences from Japan and South Korea on young Chinese people. The age modes observed for some tags in the class ``Scenario'', e.g. the growing mode of ``Driving'' and ``Travel'', and the decaying mode of ``Studying'' and ``Sports'', show differences in the situational contexts among users with different ages. In the tag class ``Emotion'', the tags that are related to melancholic moods, e.g. ``Nostalgia'', ``Lonely'' and ``Missing'', usually have a growing mode. This suggests a tendency for melancholic or negative emotions to grow with age among Chinese music listeners.

\begin{figure*}
\begin{center}
\hspace*{1.5em}
\includegraphics[width=17.9cm]{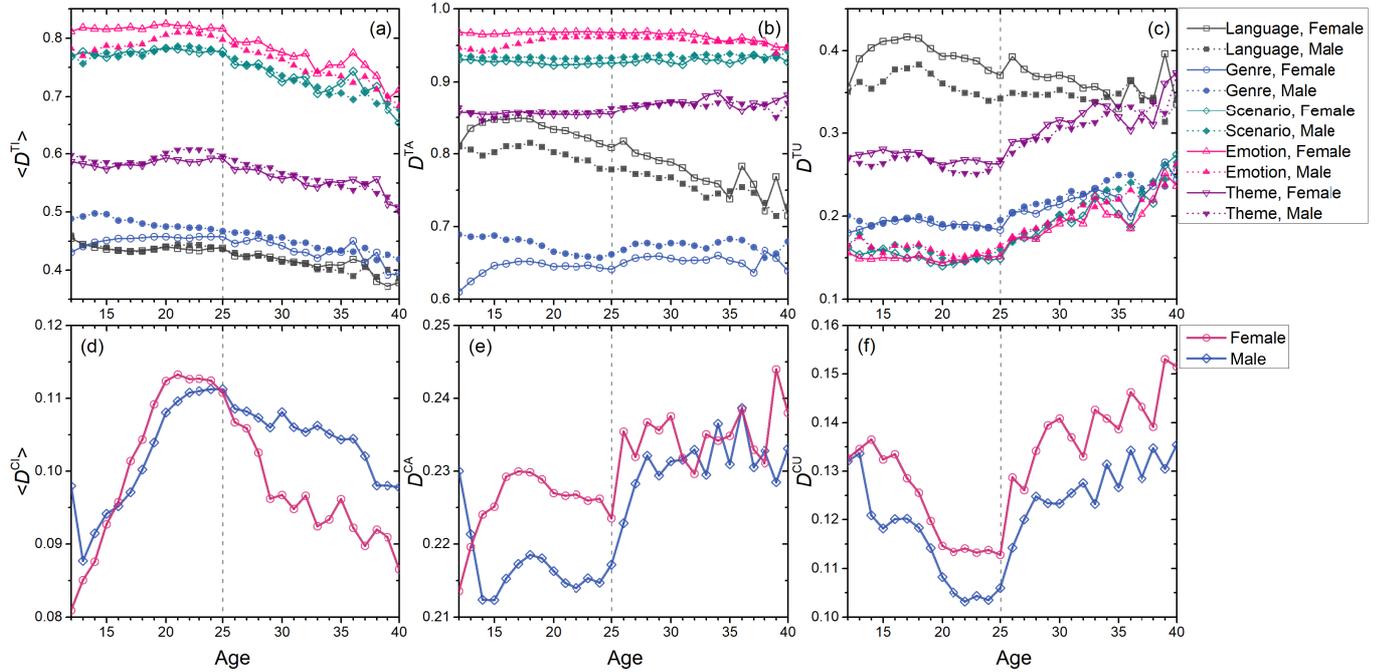}
\end{center}
\vspace{-14pt}
\caption{(Color online) Changes of the different types of diversity with age. Panels (a), (b), and (c), respectively, are the average individual tag diversity $\langle D^{TI}\rangle$, the aggregate tag diversity $D^{TA}$, and within-group tag diversity of users $D^{TU}$, for gender and different tag classes. Panels (d), (e), and (f), respectively, are the average individual community diversity $\langle D^{CI}\rangle$, the aggregate community diversity $D^{CA}$, and within-group tag diversity of users $D^{CU}$ for different ages and gender groups. The vertical dashed lines highlight the age of 25.
}\label{diversity_age}
\end{figure*}

We further investigated the relationship between age and tag diversity. For each tag class, the average individual tag diversity $\langle D^{TI}\rangle$, the aggregated tag diversity $D^{TA}$, and within-group tag diversity $D^{TU}$ of male and female users in each age group, are plotted in Figure \ref{diversity_age}(a), (b) and (c), respectively. There are significant differences between the diversity levels of various tag classes. The tag class ``Language'' has the lowest $\langle D^{TI}\rangle$ (Figure \ref{diversity_age}(a)) and the highest $D^{TU}$ (Figure \ref{diversity_age}(c)) for the majority of age ranges, indicating a strong individual heterogeneity and the largest within-group diversity of users' preferences for the language of music tracks, suggesting that a typical NCM user prefers to listen the music that is related to just a few languages, and that the language preferences within an age group are usually diversified. The tag class ``Genre'' also has a lower $\langle D^{TI}\rangle$, the lowest $D^{TA}$, and its $D^{TU}$ is also in the lower range (Figure \ref{diversity_age}(a-c)), suggesting small within-group differences in users' preferences for musical genre. On the other hand, the tag classes ``Emotion'' and ``Scenario'' have much higher $\langle D^{TI}\rangle$ and $D^{TA}$, and the lowest $D^{TU}$.

Furthermore, some tag classes show conspicuous gender differences. As shown in Figure \ref{diversity_age}(a), male users have higher $\langle D^{TI}\rangle$ for ``Genre'', possibly indicating that men have a greater openness to a variety of music preferences, and 
lower $\langle D^{TI}\rangle$ on ``Emotion'', relating to the possibility that women are more likely to respond to music in a more emotional way than men. 
Female users also have a higher within-group diversity $D^{TU}$ for the tag class ``Language'' (Figure \ref{diversity_age}(c)), especially for users under the age of 25.

In all of the three types of tag diversities, there is a remarkable transition around the age of 25, including the downward transition on $\langle D^{TI}\rangle$ for all tag classes (Figure \ref{diversity_age}(a)) and $D^{TA}$ and $D^{TU}$ for the tag class ``Language''(Figure \ref{diversity_age}(b),(c)), as well as the upward transition in $D^{TU}$ for all tag classes but ``Language'' (Figure \ref{diversity_age}(c)). Under the age of 25, users can rather 
stably maintain a higher level of individual tag diversities $\langle D^{TI}\rangle$. After the age of 25, individual diversities tend to decline, and besides the tag class ``Language'', the within-group diversities $\langle D^{TU}\rangle$ for most of the tag classes start to rise. This transition suggests that users under the age of 25 usually have a higher openness towards different types of music, and then develop stronger musical preferences for some particular music types with a growing preference bias among different users. The tag class ``Language'', the only exception to the growing trend of $\langle D^{TU}\rangle$ after the age of 25, is closely linked to the cultural influences from other countries, and its declining pattern with age potentially reflects a stronger level of multiculturality present among younger Chinese people.

Given these results, the question that naturally arises is how to interpret this transition seen around the age of 25? The following analysis on community diversities provides an insight into this question. As we have mentioned before, different from the tag information, users' music preferences cannot be reflected in the community structure if two types of tracks with a different style have almost the same followers, and the patterns of users' music preferences that can be extracted from the community structure are the cases that have driven an obvious social division. In other words, some of music tracks' social attributes can be revealed by the detection of their underlying community structure.

Specifically, we compare the three types of community diversity measures that we mentioned previously for gender and age groups. As shown in Figure \ref{diversity_age}(d), the average individual community diversity $\langle D^{CI}\rangle$ across different age groups forms an inverted U-shape-like curve for both men and women. Here, the individual community diversity increases prior to the age of 20, reaching its peak between the ages 20 and 25, and then declines after the age of 25, suggesting that listeners become less active in exploring different types of music in their early adulthood. Moreover, we can observe that this decline after the age of 25 is clearly stronger for women than for men.

At the aggregate level, the transition at the age of 25 is more obvious. As shown in Figure \ref{diversity_age}(e), there is an obvious steep rise starting around the age 25, indicating that the aggregate community diversity becomes increasingly elevated. The gender influence appears to be weaker than that observed at the individual level, whereby female users have a higher $D^{CA}$ in the ages between 14 and 30.

We further explore the within-group community diversity ($D^{CU} = D^{CA} - \langle D^{CI}\rangle$) of male and female users from different age groups. As shown in Figure \ref{diversity_age}(f), there is a remarkable valley for the age groups between 20 and 25 for both men and women, suggesting that musical preferences of different users are more homogeneous within this range of ages. There is also a steep rise between 25 and 28 years of age, which is afterwards 
turning into a stable growth trend with a rapid separation between male and female users. This difference in $D^{CU}$ is for female users generally higher than the one for their male counterparts (Figure \ref{diversity_age}(f)).

We notice in passing that the age group of 20-25 is most likely composed of college students or recently-employed college graduates, whereas the age group of 25-28 usually corresponds to the initial career-growth stage. The higher homogenization of musical tastes among college students and young professionals could therefore be attributed to the more homogeneous social structure in these environments. Thus, the transition at the age of 25 is more likely to reflect the change of the social environment of an individual, from a campus-based university life to the life of the developing career and the growing pressures to establish a family \cite{Hoganrob2004}. This interpretation fits nicely with our previously discussed observations for the age modes of genre-related tagging behavior (see Table 3), showing that at this stage of life, Chinese listeners typically develop preferences for {\it unpretentious} and {\it sophisticated} styles of music, which is also in line with the findings of Bonneville-Roussy et al. \cite{Bonneville2013}. More specifically, unpretentious music styles (e.g. pop, country, ethnic, folk) with their relaxing themes may particularly be appealing to individuals with the family life focus, whereas the artistic and creative qualities of sophisticated music (classical, jazz, blues etc.) are more attractive among adults focusing on their career and social status development \cite{Bonneville2013}.

\begin{figure}[h]
\begin{center}
\includegraphics[width=8.7cm]{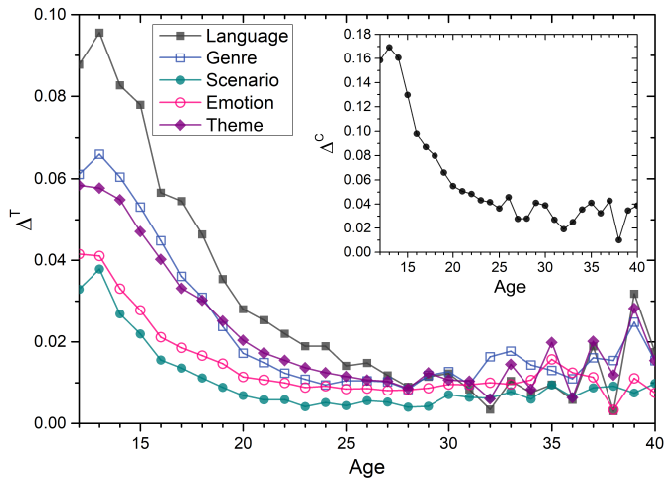}
\end{center}
\vspace{-14pt}
\caption{(Color online) The tag-based KLD ($\Delta^T$) for gender differences in each tag class as a function of age. The inset shows 
the community-based gender-differences KLD ($\Delta^C$) as a function of age.
}\label{KLD_age}
\end{figure}

Finally, tag-based and community-based gender differences in users' musical preferences and their variability with age were also investigated in our present paper. We used two KLD measures to quantify the underlying gender differences, including the tag-based KLD $\Delta^T$ for each tag class, and the community-based KLD $\Delta^C$. As shown in Figure \ref{KLD_age}, both KLD measures show highly similar trends for all investigated tag classes: They decay from a peak reached at a teen age to a more stable level attained around the age of 25. The largest gender differences for various tag classes can be observed at the age of 13, which surprisingly coincides with the age of the listeners' highest musical sensitivity (see Fig. \ref{sensitivity}). Prior to the age of 25, the tag class ``Language'' shows a much larger $\Delta^T$ relative to all other tag classes, suggesting a stronger gender-specific sensitivity to various cultures at a younger age. Besides these two KLD measures, we used the Jensen-Shannon divergence (JSD) to additionally quantify the underlying gender differences in users' musical preferences 
and their variability with age. The JSD measure is actually the symmetrized version of KLD, and as such has widely been used for the assessment of differences across distributions. This additional analysis revealed that both the tag-based and the community-based JSD measures resulted in almost identical trends as those obtained with the two KLD measures that were depicted in Fig.~11 (not shown). A summary of notations and symbols for different measures used throughout this paper is shown in Table \ref{tab_notations}.

\begin{table*}[htbp]
\scriptsize
\centering
 \caption{\label{tab_notations} Notations and descriptions of main metrics used in the paper. }
 \begin{tabular}{cm{7.5cm}m{8cm}} 
 \toprule
 Notation &  Description & Notes  \\
\hline
\rowcolor{mygray}
$\bm{A}^Y$ & The matrix describing the attention map between users with different birth years and tracks with different release years & Each column of $\bm{A}^Y$ is a normalized distribution of the release year of FP tracks for the users with the same birth year.\\
$A_i^G$ & The global distribution of user's preferences & It describes the global, average level of users' preferences for the tracks released in the $i$-th year. \\
\rowcolor{mygray}
$I_i^G$ & The average global preference that tracks in the $i$-th release years received from all users &$I_i^G = A_i^G / \rho_i$, where $\rho_i$ is the ratio of the number of tracks released in the $i$-th year.\\
$\bm{R}$ & The matrix describing the relative attention that tracks in the $i$-th release years received from users with a birth year $j$ & Its element: $R_{ij} = A^Y_{ij}/A_i^G$\\
\rowcolor{mygray}
$S_{j}(i-j)$ & User's sensitivity to music with birth year $j$ for the tracks released in year $i$  &  $S_{j}(i-j) = R_{ij}\langle R_{kj}\rangle^{-1}$\\
$D_{Gi}^{TI}$ & The $i$-th user's individual tag diversity within the tag class $G$ in a certain demographic group &It measures the degree of homogeneity of a user's preference for a given music tag in a given tag class. \\
\rowcolor{mygray}
$D_G^{TA}$ & The aggregate tag diversity of the tag class $G$ in a certain demographic group &  It measures tag diversity of a group's users in an aggregate fashion. \\
$D_G^{TU}$ & The within-group tag diversity of the tag class $G$ in a certain demographic group &  $D_G^{TU} = D_G^{TA} - \langle D_G^{TI}\rangle$, which measures the tag differences in a given tag class among users belonging to the same demographic group.  \\
\rowcolor{mygray}
$D_i^{CI}$ & The individual community diversity of the $i$-th user in a certain demographic group & It measures the degree of homogeneity with which user's FP tracks are distributed across different communities. \\
$D^{CA}$ & The aggregate community diversity of a certain demographic group & It measures the community diversity of a group's users in an aggregate fashion. \\
\rowcolor{mygray}
$D^{CU}$ & The within-group community diversity of a certain demographic group & $D^{CU} = D^{CA} - \langle D^{CI}\rangle$, which measures the diversity of user communities belonging to the same demographic group.   \\
$\Delta_G^T$ & The tag-based Kullback-Leibler divergence  &  It measures the gender differences in the attention distribution of users in a certain demographic group for the tag class $G$. \\
\rowcolor{mygray}
$\Delta^C$ & The community-based Kullback-Leibler divergence  &  It measures the gender differences in the community distribution of user's FP tracks in a certain demographic group.  \\
  \bottomrule
 \end{tabular}
\end{table*}

\subsection{Relationship between music preferences and regional economic development}

\begin{figure*}
\begin{center}
\includegraphics[width=17.9cm]{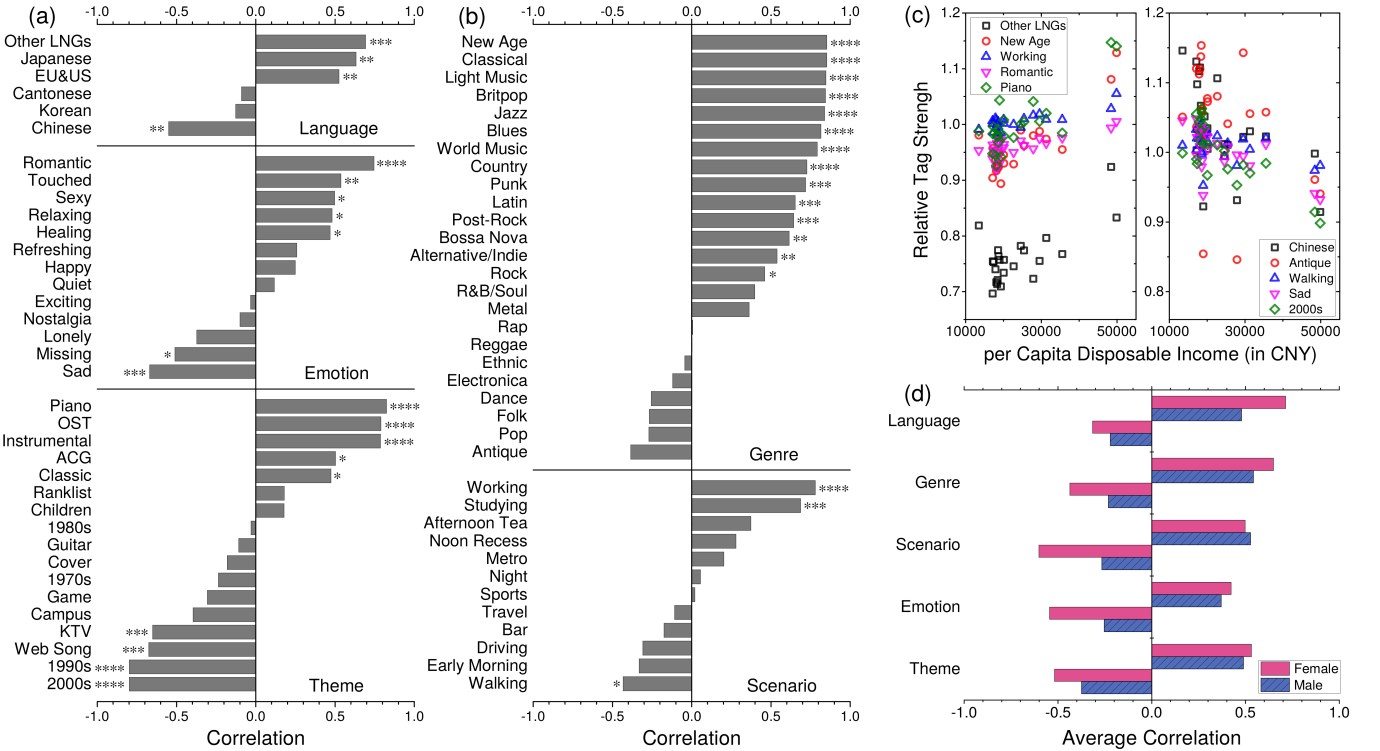}
\end{center}
\vspace{-14pt}
\caption{(Color online) The correlations between users' music preferences and regional economic development. (a) and (b) The correlation between the aggregated tag strength and the per capita disposable income at the province level for each tag. The tags are ranked along the decaying order of the correlation for each tag class. (c) The correlation patterns for the tags with the strongest correlation in each tag class. The left panel shows the strongest positive correlations, and the right one shows the strongest negative correlations. (d) The average correlation coefficients of each tag class for female and male users. The coefficients for positive correlations and negative correlations are averaged, and exclude the tags with opposite correlations for male and female users.
}\label{tags_corr}
\end{figure*}

At both individual and societal levels, the previous research has shown a strong connection between cultural preferences and economic 
development \cite{Mellander2018}. Eagle and colleagues \cite{Eagle2010} have discovered a remarkable association between the diversity of people's social connections and the regional economic development level of their location.

Since users' NCM information does not reveal their economic status, we attempted to detect such an association at the societal level. From the yearbooks released by the National Bureau of Statistics of China and the provincial bureaus of statistics, we collected regional economic indicator data (including GDP per capita, per capita disposable income, and the output value of each type of industry) of all 31 provinces (including all four municipalities and five autonomous regions) in mainland China in 2015. 

Given that the population of China is dominantly Han Chinese, and some provinces have a higher proportion of minority groups than others, the pertinent differences in musical preferences across various provinces could additionally be affected by ethnicity-related factors, besides age, gender, or economic development. We therefore excluded from further analysis all provinces with an ethnic minority proportion above 20\% (including five autonomous regions: Inner Mongolia, Guangxi, Tibet, Ningxia, Xinjiang; and the three provinces: Yunnan, Guizhou, Qinghai). We then focused our subsequent analyses on the rest 
of 23 provinces in mainland China, that included a total of 249 different cities.

We treated users in each province as representatives of a group and we first obtained the tag vector-set for each such group. We then calculated the Pearson correlation coefficients between each tag's strength and the indicator of economic development at the province level. Several economic indicators, including GDP per capita, per capita disposable income, and per capita tertiary industry output value, were entered into the analysis. The per capita disposable income yielded the strongest correlations for most of the investigated tags. In our following discussions, we thus employ per capita disposable income as the representative indicator of economic development.

Figure \ref{tags_corr}(a) and (b) shows the correlations between each tag and the per capita disposable income. We see that the calculated correlations range from strongly positive to strongly negative, indicating the existence of the relationship between economic development and users' preferences for some tags. The cases with the strongest positive and strongest negative correlations for each tag class are additionally plotted in Figure \ref{tags_corr}(c). It is noteworthy that, in the tag class ``Emotion", most tags with positive emotional connotations also have positive correlations with economic development, whereas tags with melancholic or negative emotional connotations, such as ``Lonely", ``Missing" and ``Sad", show negative correlation with economic indicators. We note that this finding could also serve as a novel regional-level evidence for the relationship between economic indicators and individual well-being \cite{Zagorski2010}, as reflected via musically associated moods and emotions \cite{Parkinpress} in different regions in China. Moreover, for the tag class ``Language", we find that the tag ``Other LNGs" (which stands for other languages), yields the strongest positive correlation with economic development. Thus, a higher economic status can be associated with more opportunities for the growth of niche cultures and subcultures in China.

When considering gender differences in our correlation analyses, we found that on average, the tag strength of female users is stronger related to the level of economic development than that of male listeners of NCM. In Figure \ref{tags_corr}(d), we see the average values of the Pearson correlation coefficients for each tag class split by gender (separately for positive and negative correlations, respectively). We see that for the majority of tag classes, these correlations (both positive and negative) are higher for female users than for males, indicating that women's musical preferences in China are more strongly tied to economic factors.

\begin{figure*}
\begin{center}
\includegraphics[width=13cm]{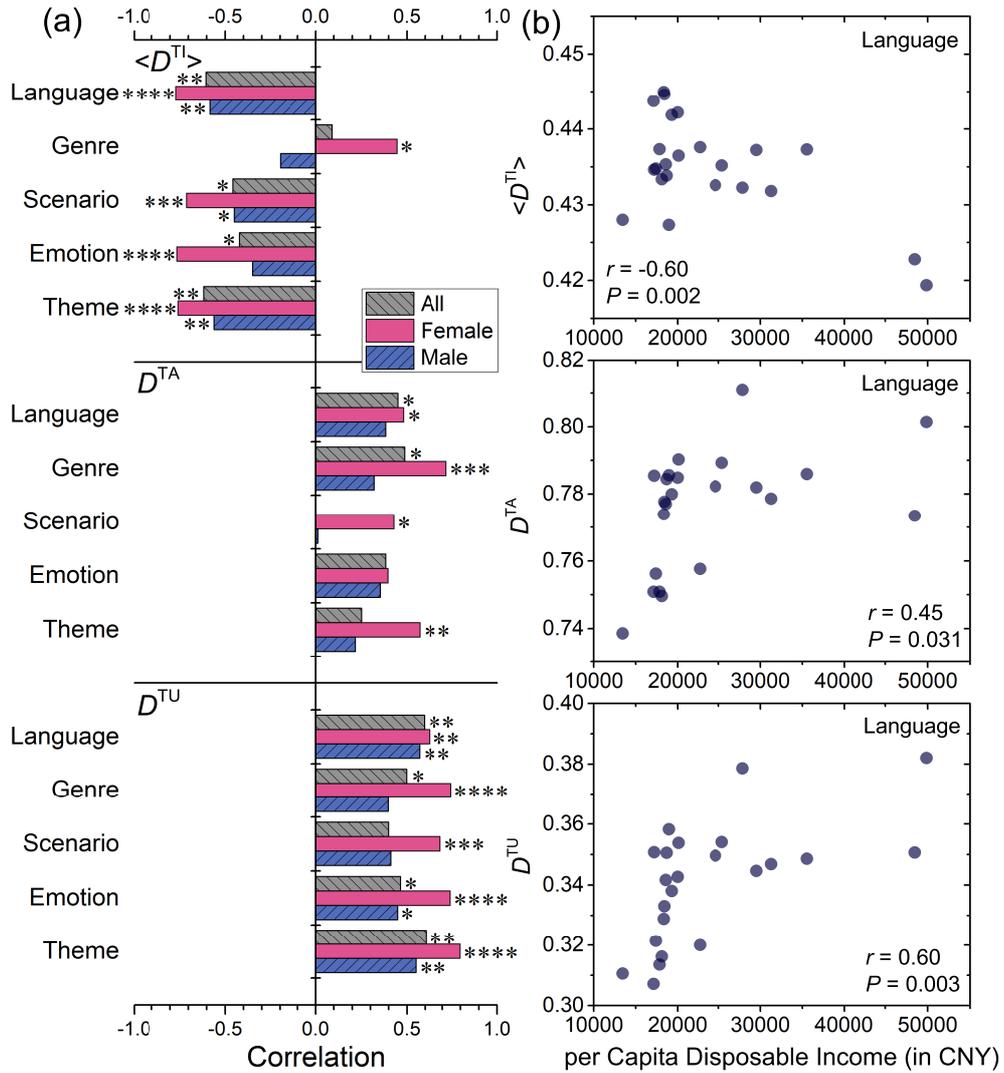}
\end{center}
\vspace{-14pt}
\caption{(Color online) (a) The correlations between three types of tag diversities $\langle D^{TI}\rangle$, $D^{TA}$, and $(D^{TA}-\langle D^{TI}\rangle)$ and the per capita disposable income at the province level for each tag class. Gray, pink and blue bars, respectively, correspond to the cases of all users, female users, and male users. (b) The scatter plots with the correlations between tag diversities $\langle D^{TI}\rangle$, $D^{TA}$, and $D^{TU}$ of tag class ``Language'' and the per capita disposable income at the province level. The currency unit of the per capita disposable income was CNY.
}\label{DT_corr}
\end{figure*}

We further investigated the relationship between various tag diversity measures and regional economic development. Figure \ref{DT_corr} shows the correlations between the three tag diversity measures and the per capita disposable income at the provincial level (Fig. \ref{DT_corr}(a)), and the typical patterns of these relationships for the tag class ``Language'' (Fig. \ref{DT_corr}(b)). Except for ``Genre'', the average individual tag diversities ($\langle D^{TI}\rangle$) of all other tag classes are negatively correlated with economic development. This may suggest that users in developed provinces are more likely to prefer only a few specific kinds of music. On the other hand, the aggregated tag diversity ($D^{TA}$) and the within-group tag diversity ($D^{TU}$) consistently show positive correlations with economic development across different tag classes. These result patterns may further suggest that economic development drives the diversification of musical preferences in the whole population, but not at the individual level. Moreover, similar to the relationship between economic indicators and tag strengths, these correlations between various tag diversity measures and economic development are consistently higher for the female than the male subpopulation of users.

\begin{figure*}
\begin{center}
\includegraphics[width=14.7cm]{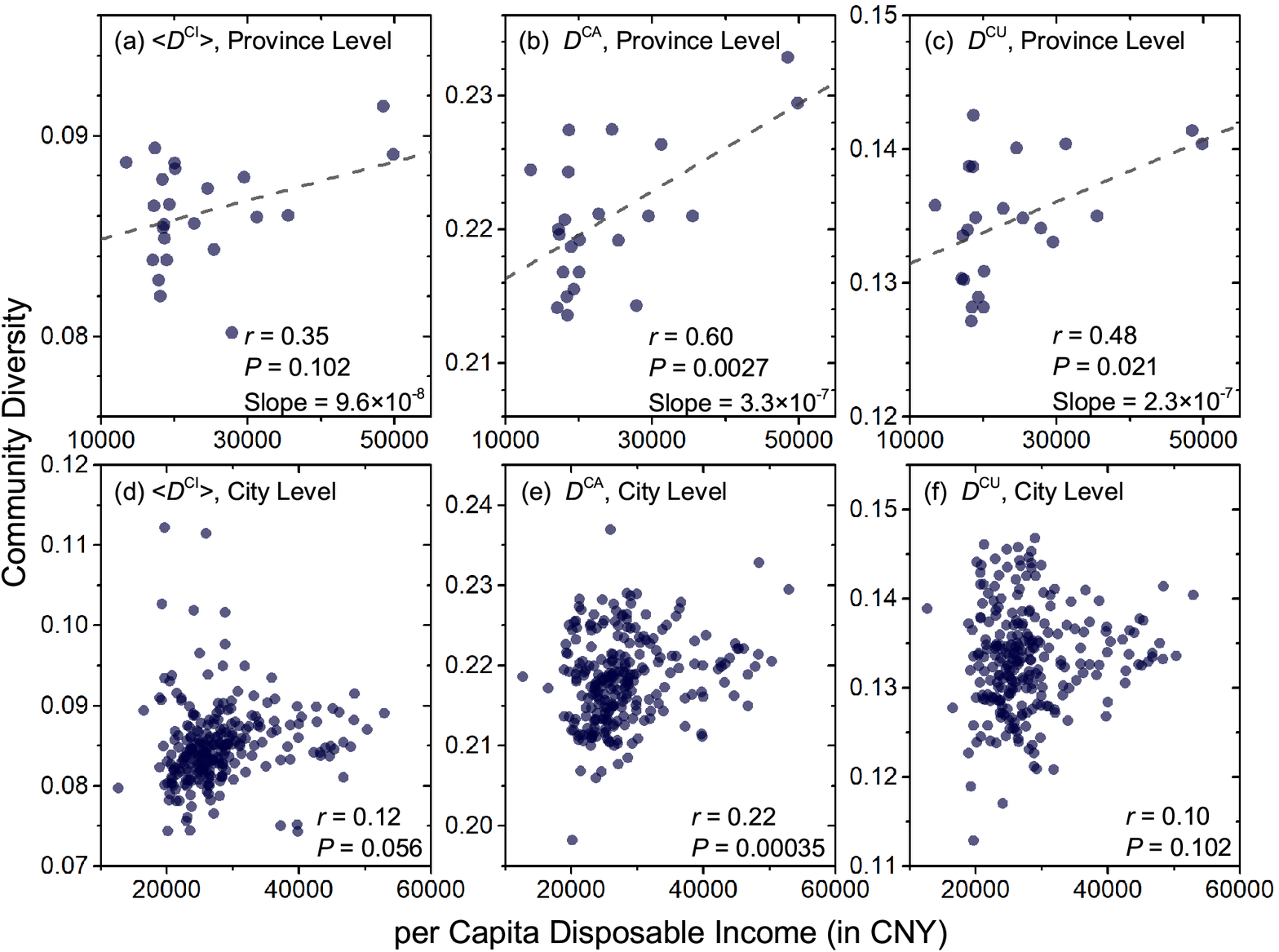}
\end{center}
\vspace{-14pt}
\caption{(Color online) The relationship between the average individual community diversity $\langle D^{CI}\rangle$ (panel (a)), the aggregated community diversity $D^{CA}$ (panel (b)), and the within-group community diversity $D^{CU}$ (panel (c)) and the per capita disposable income at the province level. Panels (d), (e), and (f), respectively, show the relationships between each diversity measure and the per capita disposable income at the city level. The currency unit of the per capita disposable income was CNY.
}\label{economic}
\end{figure*}

However, the average individual community diversity ($\langle D_i^{CI}\rangle$) has the opposite patterns of relationship with regional economic development. For example, as shown in Figure \ref{economic}(a), $\langle D^{CI}\rangle$ seems to exhibit a weak positive correlation with the per capita disposable income at the provincial level; however, this relationship was statistically not significant ($p>0.05$). Actually, even a significant correlation would here not contradict the cases of the average individual tag diversity $\langle D_i^{TI}\rangle$, but instead it would paint a picture in which users in developed provinces would have narrower musical preferences but wider links to other communities, since the detected large communities are all tag-crossed.

The aggregate community diversity $D^{CA}$ had a significant positive association ($r = .60$, $p=.0027$) with regional economic development (Figure \ref{economic}(b)), suggesting a rather strong relationship between economic development and users' musical preferences at the aggregate level. The within-group diversity $D^{CU}$, as measured by ($D^{CA} - \langle D_i^{CI}\rangle$), also positively correlated ($r= .48$, $p=.021$) with the regional economic development (Figure \ref{economic}(c)). Since the correlation between $D^{CU}$ and economic development is greater than that between $\langle D^{CI}\rangle$ and economic indicators, it is easy to find that the diversity of users seems to play here the major role and that economic development is mainly associated with the growing diversification in the population of music listeners.

In comparison to the province-level correlations, the relationship between community diversity and regional economic development at the city level was generally much weaker. Here, the measured correlations between regional economic development and $\langle D^{CI}\rangle$, $D^{CA}$, and $D^{CU}$ measures were $r= .12$ ($p=.056$), $r= .22$ ($p=.00035$), and $r= .10$ ($p=.102$), respectively (see Figure \ref{economic}(d), (e), and (f)). As shown in Figure \ref{economic}(d), the data spread around a Y-shaped region with several cities having low per capita disposable income but high average community diversity ($\langle D^{CI}\rangle$). The patterns of $D^{CU}$ at the province level and $D^{CA}$ at the city level in Figure \ref{economic}(c) and (e) suggest 
a more intricate relationship between economic development and musical preferences, that would thus require a further study across different societal levels and at various scales.

\begin{figure*}
\begin{center}
\includegraphics[width=12.8cm]{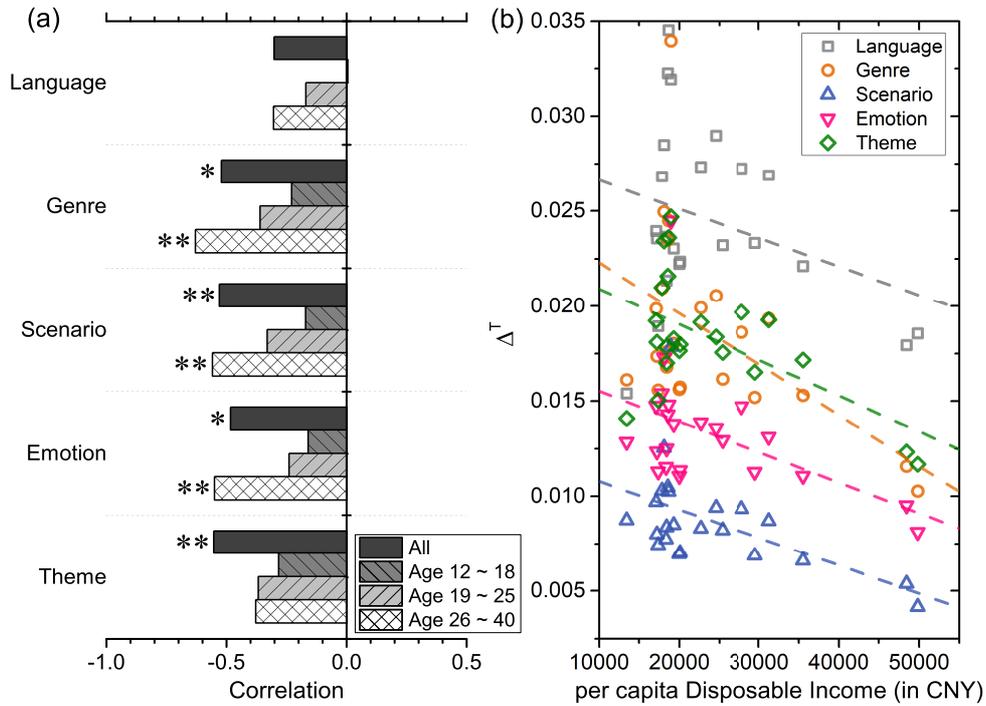}
\end{center}
\vspace{-14pt}
\caption{(Color online)
(a) The correlations between the gender differences in musical preferences (measured by the tag-based KLD $\Delta^T$) and the disposal income of each province for different tag classes. Correlations are shown for all users, the users aged between 12 and 18, 19 and 25, and 26-40. (b) Gender differences ($\Delta^T$) between users in each province as a function of the per capita disposable income of each province. The currency unit of the per capita disposable income was CNY.
}\label{KLD_corr}
\end{figure*}

Similar correlations were also found between gender differences and regional economic development. We calculated gender differences for users in each province using the tag-based KLD $\Delta^T$ for different age groups.
As shown in Figure \ref{KLD_corr}(a) and (b), for all tag classes, the tag-based KLD $\Delta^T$ correlated negatively with the disposal income of each province. Thus, the gender gap in users' musical preferences decreased with an increasing provincial economic development. This finding suggests that 
economic development could thus contribute to the reduced gender gaps in China when it comes to development of musical tastes. Interestingly, this negative correlation between gender differences and income is stronger in the older age group (26-40) than in younger age groups (Fig. 15(a)), even though the observed gender differences generally tend to decrease with an increasing listeners' age (Figure \ref{KLD_age}). In addition, we note that the outcomes of the correlation analyses shown in Fig.~15 were almost identical when using the Jensen-Shannon divergence (JSD) instead of the KLD measure (not shown). 

Similar to the study of Mellander et al. \cite{Mellander2018}, we also found strong positive correlations between income and ``sophisticated" types of music (e.g. jazz, classical, world), but we did not observe such a relationship between economic factors and what Mellander et al. \cite{Mellander2018} called the contemporary music types (i.e. rap, soul, and reggae). This is an interesting cultural difference between US American and Chinese music listeners, suggesting that while we can still argue for the existence of cross-cultural universals in the development of musical preferences, the relationship between economic factors such as income and tastes for different musical genres is rather culture-dependent. Thus, various cultures may simply differ in how they define ``contemporary" music styles. For example, in the context of the social identity theory and its application to musical preferences 
development \cite{tekman2002}, people build their musical preferences jointly with others with whom they share similar or same values, and these values are often reflected in shared interests in similar music types. So, obviously, values represented by American rap or reggae music are less shared among Chinese listeners, and as a result, different from what Mellander and colleagues \cite{Mellander2018} observed for their US American sample, we did not 
find any correlations for these types of music in our NCM dataset with Chinese music listeners.

\subsection{Novel methodological features of our study}

With respect to methodological issues, one of the greatest challenges in the study of musical preferences are the constraints set by genre-related approaches \cite{Greenberg2015}. Here, self-rated musical preferences are given to a list of items characterized by {\it a priori} specified musical genres. However, as has often been pointed out, genres are only made-up labels that have been invented by the music industry to broadly categorize music \cite{Greenberg2015} and as such are rather ill-defined for describing individual musical preferences \cite{Rent2011,lambiotteausloos}. In our present study, we have tried to circumvent this problem by using the genre information that is actually constructed by collaborative tagging of NCM users. Instead of using predefined genres for each considered track in our dataset, the genre information was assigned by the users themselves via their tagging behavior. Such information is interesting because it contains individual perceptions of psychological and sonic features of the considered musical genres. Thus, the ultimate number of genres and their distribution across different tracks was in our study defined by the NCM users. 

This approach is both innovative and relevant as there is currently no general consensus about which and how many particular genre types should be studied when investigating musical preferences \cite{Rent2011}. Besides, labeling music tracks by a certain genre type can sometimes pose a nontrivial challenge for music experts as well (e.g. if a given track contains a mixture of different genres), so that genre information originating from aggregate tagging behavior can yield a more representative consensus about the various disputed cases. Of course, concomitantly, such behavior could also introduce some uncertainty into the actual correspondence between the assigned tags and the actual genre of a given track. However, it is rather difficult to estimate the magnitude of this effect, given that quite often, only a few expert users can rather quickly and accurately set the tag of a track. 

Furthermore, as pointed out by Bonneville-Roussy and colleagues \cite{Bonneville2013}, earlier studies addressing the variability of musical preferences with age employed only pop-music tracks (see e.g., \cite{Holbrook1989}). Our present study instead analyzed differences in preference behavior for a much wider range of musical styles and situations in which music can be enjoyed, allowing us thus to address the question of whether age trends in musical preferences vary by musical genre and context. While our genre sample is arguably sufficiently representative of the Chinese musical tastes spectrum, we think that even more diverse selections of music, including those rated by different attributes than genre, should be used in future studies.

The limitations of user sample sizes and narrow age ranges of music listeners that pervade earlier investigations of musical preferences were overcome in our study by drawing a large sample of more than three million users distributed across 29 different age groups (ranging from the age 12 to 40). For this range of listeners' ages, on which our analyses mostly focused, our sample included at least 846 users in each individual age group. Remarkably, our sample had over 5,000 users in 18 out of 29 investigated age groups, and in seven groups we had over 30,000 individuals, with the maximum number at the age of 21 (a total of 49,870 users with 29,464 males, 16,597 females, and 3,809 gender-unknown users). To the best of our knowledge, this is by far probably the largest sample ever used in a study of musical preferences, and certainly the largest one for the population of Chinese music listeners. The statistical power yielded by such a large sample size warrants a reliable detection of curvilinear trends that may otherwise remain uncovered in smaller 
samples (for a related discussion, see Ref. \cite{Bonneville2013}).

\section{General discussion and future research directions}

The advent of online music-based social networks and the increasing accessibility of data on virtual behaviors have enabled unprecedented large-scale studies of human cultural dynamics. Our present study investigated the formation of musical preferences and their variability with demographic factors by analyzing 30,562,590 music playlists, 2,247,960 songs, and behaviors of 3,028,351 users of the NCM online music-based community, one of the largest online music platforms in China.

Combining the tools from information sciences, statistical physics, and complex networks theory within the context of Big Data analysis, we 
unveiled several systematic patterns in the formation and evolution of musical preferences in China. We further employed a series of statistical measures to quantitatively characterize different variables related to both individual-level and global musical preferences in the studied population, including users' tagging diversity, community diversity, individual and global musical preferences, the decay pattern of music influence, and user's sensitivity to music. We then modeled the obtained empirical data distributions identifying various features of musical preferences, including peak sensitivity to popular music that was reached at the age of 13, thereby strikingly matching recent findings observed for US American users of Spotify, another online music-based community \cite{Stephens-Davidowitz2018}. We finally performed a series of correlational analyses to address the relationships between different music preference variables and demographic factors such as age, gender, and economic development.

The striking similarity with recent online music community analyses of non-Chinese music listeners (e.g. \cite{Stephens-Davidowitz2018}) allows us to assume some level of cross-cultural universality in the formation of musical preferences, but not in the relationship between economic development and genre-based musical tastes (cf. \cite{Mellander2018}). By examining the features of the user-music bipartite network, we observed a stable temporal trend of musical tastes among Chinese listeners of NCM: The musical preferences decay slowly in a power-law-like fashion. Such specific patterns have been identified in some online social-media behaviors, e.g. in the outbreak of Internet memes and in the spreading of popular topics \cite{Crane2008,Leskovec2009}.

However, the observed decay of musical preferences in our study was much slower than what has previously been reported for other online behaviors. Namely, the complete decay process in our present study spreads over the extended period of 20 years, suggesting a remarkably stretched phase of musical taste formation and new music discovery, accompanied by a slow taste freezing in the population of Chinese music listeners. In other words, the change of music preferences is often a generational phenomenon in China. This interesting finding raises the question of whether the characteristics of musical preferences 
are generalizable to other types of nonmusical preference behaviors, such as those for films, dance, fashion, etc. Further comparative studies across different cultural products will be necessary to address this more global level of potential universality of cultural preferences.

We further mapped the age trends of each tag and classified them into nine basic age modes. In terms of individual musical sensitivity, we found that teenagers are more sensitive to music than adults, which is consistent with recent Big Data analyses (e.g. \cite{Stephens-Davidowitz2018}). After the age of 25, the individual preference diversities for both tags and communities start to decline, and the within-group diversities rapidly increase for most of the investigated cases. In other words, individual music tastes are greatly shaped by the musical environments and events occurring in the teen ages (especially around the age of 13), and at the time of the early adulthood (ages 20-25), representing the two major transitions in the formation of musical preferences. Individuals then tend to further explore different types of music but also to develop more stable and personalized styles following the age of 25. 

However, with respect to these critical periods and major transitions in the development of music preference behaviors, our analyses reveal a remarkable discrepancy between the conclusions of earlier offline (e.g. \cite{Holbrook1989}) and more recent online studies of musical 
preferences (e.g. \cite{Stephens-Davidowitz2018}). More specifically, previous offline studies, that were also typically limited to small sample sizes, found an inverted U-shaped pattern characterizing the development of musical preferences, with a peak observed around the age of 24 \cite{Holbrook1989}. On the other hand, large-scale studies of preference behaviors in online music-based communities \cite{Clauset,Stephens-Davidowitz2018}, including our present investigation, suggest that the period of maximum sensitivity to music starts much before the early adulthood, in the early teens of music listeners. Since these differences may be due to many factors (e.g. differences in age-ranges of participants, differences in the number of 
investigated music tracks and genres etc.), it will remain an evident challenge for future research to identify the underlying origin of this discrepancy.

Critical periods in the development of musical preferences reported in our study are well supported by interactionist 
theories (e.g. \cite{Swann2002}), postulating that musical preferences emerge and develop in accordance with listeners' basic psychological needs and emotions, their reinforcement, and as the reflection of the emerging social identities that can be traced across different developmental phases. For example, the valley of the within-group community diversity that was observed in our study for the age period 20-25 corresponds largely to the more homogeneous and stable phase of the university campus life, whereas the rapidly growing within-group diversities after the age of 25 are more likely to be associated with the later social influences prevalent in the emerging professional careers. 

Our study has also revealed strong gender differences in the development of musical tastes. In general, the within-group differences in musical preferences of women are larger than those of men, even though their genre-related individual diversity is lower than in male users. These findings largely break through the individual perspectives reported in previous studies \cite{Bonneville2017a}, 
but they also expand our understanding of musical taste formation towards the multidimensional, socio-psychological-technological perspective, 
revealing an intrinsic complexity of gender trends in online music preference behaviors. 

Importantly, we observed the strongest gender differences among those online music listeners who were in their teens (see Fig. 11). This is in contrast with a previous study of US users of Spotify (e.g. \cite{Kalia2015}), in which male and female teenagers were found to share highly similar listening behaviors. This discrepancy between adolescent Chinese and US music listeners suggests that unlike age, the role that gender plays in music preference formation is rather culture-specific. However, since the explorative study of Kalia \cite{Kalia2015} did not consider a detailed structure of musical preferences and largely focused on music popularity, further comparative and more comprehensive analyses will be necessary to draw conclusive statements about cross-cultural gender differences in the formation of musical preferences. 

Our study also identified a relationship between musical preferences and regional economic development. We found that economic development is mainly related to within-group diversity in music preferences, but not to diversity at an individual level, suggesting thus the growth of some minority music circles in China. The negative correlations between economic indicators and tag strengths for melancholic/negative emotions suggest a possible positive influence of economic development on listeners' well-being and life outlook. Moreover, the generally stronger correlations in female users indicate that the musical preferences of women could be more sensitive to economic development. 

Interestingly enough, the observed negative correlation between gender differences in music preferences and economic indicators in our study (Fig.~15(b)) is in contrast with a recently observed positive relationship between gender differences in preference behaviors and economic 
development \cite{falkhermle}, suggesting that these gender differences can actually widen with a more egalitarian availability of resources and the associated enhanced formation of gender-specific preferences. However, we note here that this positive relationship between gender differences in preferences and economic factors \cite{falkhermle} was not observed for musical preferences but only for behaviors that involved economic decision making processes, e.g. for preferences in risk-taking or in various types of costly cooperative behaviors. It is therefore possible, as our current study seems to suggest, that preference behaviors which do not involve economic decisions (such as preferences for different musical genres) tend to show a negative correlation with economic development, because gender differences in non-costly behaviors are more likely to diminish in the presence of a gender-equal access to resources. The observed decline in gender differences in music preferences with an increasing economic development in our present paper is supported by the 
social role theory \cite{Eagly2018}, according to which gender differences in preference behaviors should weaken in more developed regions, where higher economic development and the associated declining relevance of conventional gender roles are expected to narrow gender gaps in preference behaviors. 

In spite of its largely explorative nature and a variety of methodological challenges, Big Data technology has a wide ranging applicability in a number of domains \cite{wangfujita}, where it can successfully complement various predictive modeling techniques. One potential area of application of our current study is music recommendation \cite{skleeetal2010} in online music streaming services and music-based online communities such as the NCM or Spotify. For example, similarly to a previous proposal \cite{Rent2011}, the information obtained in our study from user-based genre-tagging of music tracks and from other online behaviors may help in capturing the latent structure of users' musical preferences much better than the usually employed procedures with artificially labeled music genres. Future studies should therefore assess the capacity of our analyses for predicting individual online preference behaviors, and evaluate their potential usage in online music recommendation systems.

Datasets collected in our study could also be employed to calibrate future large-scale agent-based models \cite{Bonabeau2002,hadzicuiwu}, that would then simulate various aspects of users' preference behaviors which are otherwise hidden or difficult to extract from music-based social networks. For example, relative contributions of different intervening variables such as factors related to online technology use or social influence exerted among users could be estimated by using agent-based simulations. Such models could further investigate how attention diffusion \cite{Candia2019}, opinion flow \cite{Lichten2016,weisu}, or other spreading processes \cite{wuahadzi} in online social networks shape the formation of both individual and global musical preferences. Due to their {\it generative} character, agent-based models would enable us to understand not only what types of global patterns can possibly exist in online music-based communities, but more importantly, they could potentially inform us about how these patterns actually emerge and evolve from multiple local preference behaviors and interactions occurring among millions of connected individuals. 

In the meantime, while such ambitious large-scale agent-based modeling is still in progress, next studies should go beyond mere correlational analyses and address potential interaction effects between age, gender, and economic development, e.g. via analyses within the context of the structural equation modeling framework (see e.g. \cite{Bonneville2013}) or linear mixed effects models \cite{bates}. 

Finally, it could also be highly useful for the future playlist generation and recommendation system design to investigate if any changes in the patterns of musical preferences can be observed in the ongoing post-COVID-19 era relative to the previously identified, pre-COVID patterns of musical tastes that have been reported in our present paper. Indeed, in the wake of the ongoing pandemic, music has served the role of a connecting tissue \cite{greenbergdecety} bringing together millions of people in times when direct physical contact was rare and explicitly discouraged as an integral part of epidemic mitigation strategies. It would therefore be promising to investigate if preference behaviors and their relationship to demographic indicators might have changed since the outbreak, and moreover, which dynamic trends can be observed over the whole time course of the pandemic. 

\section{Conclusions} 

This study presents a Big Data analysis of global patterns in musical taste formation in China based on real behavioral data obtained from millions of social media users. Employing advanced methods from information sciences, statistical physics, and network theory, we discovered universal age and gender trends in musical preferences formation, as well as global patterns of musical sensitivity and genre-based diversity in musical tastes.

Our analyses revealed two major transitions in the development of musical preferences: One is associated with the peak sensitivity to music influence (occurring around the age of 13), and the other is related to musical taste freezing, i.e. the decaying trend in the formation of musical preferences (around the age of 25). Similar to earlier studies with participants from US and UK, we observed that whilst the interest in some musical genres and listening contexts decreased in adulthood of Chinese music listeners, their preferences for other musical dimensions actually increased with age. 

Further comparisons with recent analyses of online preference behaviors of Western listeners allowed us to surmise that these features in the development of music preferences may be universal, shared across many different cultures and times. These findings lend support to interactionist theories \cite{Swann2002} that highlight the importance of individual-environment relationships and emphasize preferences for those musical surroundings that reinforce the traits of an individual's personality, emotions, and values, and reflect the emerging social identity of an individual. 

In addition to universal properties in musical taste formation, we identified some culture-specific features including eight distinct communities of music listeners in the population of NCM users, and a relationship between regional economic development and preferences for specific music genres in China which has not been reported previously for Western music listeners. Specifically, the correlation patterns typically observed in US music listeners between some musical genres (e.g. rap, soul, or reggae) and economic indicators were not found in the Chinese NCM population of listeners, as the values represented by these genres are, in accordance with predictions of the social identity theory \cite{tekman2002}, generally less shared among Asian music consumers. We also found that in their adolescence, relatively to their male counterparts, female users of music-based social media in China were characterized by more heterogeneous music preference behaviors that were also stronger associated with economic factors. This gender gap in the adolescence was observed to decrease with the onset of the early adulthood and with the increasing economic development.

Our findings thus not only provide the large-scale empirical evidence for the existence of universal patterns in musical taste development, but they also reveal a rich and novel set of features characterizing the emergence of culture-dependent preference behaviors, that can provide further insights into regional cultural diffusion and music popularization processes. Moreover, our findings reveal remarkable differences between earlier offline and more recent online studies of musical preferences, that may be due to differences in the studied sample sizes but also due to the ever-growing availability of music and the increasing sophistication of its use in online networking technologies. Importantly, our study has shown that applying the network theoretic formalism to large-scale behavioral datasets can serve as a powerful tool for analyzing the community structure and the diversity of cultural links in online music platforms such as the NCM, enabling us to uncover musical preference trends and their global features that otherwise would have remained undetected.

These findings contribute to our multidisciplinary understanding of musical preference behaviors in online social networks, with implications for a wide variety of applied domains such as assessment of online content popularity, user profiling, community detection in online music platforms, and the design of recommendation systems for online social media. Taken together, we hope that our present results, and the studies that will follow, will help us lay the foundation for a more general, quantitative, and integrative theory of musical preferences formation, that will advance our understanding of universal properties of human culture and its evolutionary dynamics.

\section*{Acknowledgments}
This work was supported by the National Natural Science Foundation of China (grant no. 62073112), the Centre for Communication Research at City University of Hong Kong (grant no. 9360120), and the Hong Kong Research Grants Council (grant no. 11505119).



%
%
%
\bibliographystyle{elsarticle-num}

\end{document}